%% file: n1052_midi.tex
\documentclass[a4paper,fleqn,usenatbib]{mnras}

\usepackage[british]{babel} 
\usepackage{newtxtext}
\usepackage[frenchmath,varg]{newtxmath}

\usepackage[T1]{fontenc}
\usepackage{graphicx}	
\usepackage{amsmath}	
\usepackage{amssymb}	
\usepackage[dvipsnames]{xcolor}
\hypersetup{breaklinks=true,colorlinks=true,linkcolor=NavyBlue!50!Emerald,citecolor=NavyBlue!50!Emerald,filecolor=blue,urlcolor=Gray} 

\title[A Compact Jet in NGC 1052]{A Compact Jet at the Infrared Heart of the Prototypical Low-Luminosity AGN in NGC 1052}

\author[J.\,A. Fern\'andez-Ontiveros et al.]{
  J.A. Fern\'andez-Ontiveros$^{1,2,3,4}$\thanks{E-mail: \textsf{\href{mailto:j.a.fernandez.ontiveros@gmail.com}{\color{Black}j.a.fernandez.ontiveros@gmail.com}, \href{mailto:juan.fernandez@inaf.it}{\color{Black}juan.fernandez@inaf.it}}}, N. L\'opez-Gonzaga$^{5}$, M.A. Prieto$^{1,2}$, J.A. Acosta-Pulido$^{1,2}$, \newauthor
E. Lopez-Rodriguez$^{6}$, D. Asmus$^{7}$, K.R.W. Tristram$^8$
\vspace{0.2cm}\\
$^1$Instituto de Astrof\'isica de Canarias (IAC), C/V\'ia L\'actea s/n, E--38205 La Laguna, Tenerife, Spain\\
$^2$Universidad de La Laguna (ULL), Dpto. Astrof\'isica, Avd. Astrof\'isico Fco. S\'anchez s/n, E--38206 La Laguna, Tenerife, Spain\\
$^3$Istituto di Astrofisica e Planetologia Spaziali (INAF--IAPS), Via Fosso del Cavaliere 100, I--00133 Roma, Italy\\
$^4$National Observatory of Athens (NOA), Institute for Astronomy, Astrophysics, Space Applications and Remote Sensing (IAASARS), GR--15236, Greece\\
$^5$Leiden Observatory, Leiden University, PO Box 9513, 2300 RA Leiden, The Netherlands\\
$^6$SOFIA Science Center, NASA Ames Research Center, Moffett Field, CA 94035, USA.\\
$^7$School of Physics \& Astronomy, University of Southampton, Southampton SO17 1BJ, UK\\
$^8$European Southern Observatory, Alonso de C\'ordova 3107, Vitacura, Santiago, Chile
}

\date{Accepted: 2019 March 08; Revised: 2019 March 25; Received 2018 December 19}

\pubyear{2019}

\begin{document}
\label{firstpage}
\pagerange{\pageref{firstpage}--\pageref{lastpage}}
\maketitle

\graphicspath{ {./figs/} }

\begin{abstract}
The feeble radiative efficiency characteristic of Low-Luminosity Active Galactic Nuclei (LLAGN) is ascribed to a sub-Eddington accretion rate, typically at $\log(L_{\rm bol}/L_{\rm edd}) \lesssim -3$. At the finest angular resolutions that are attainable nowadays using mid-infrared (mid-IR) interferometry, the prototypical LLAGN in NGC\,1052 remains unresolved down to $< 5\, \rm{mas}$ ($0.5\, \rm{pc}$). This is in line with non-thermal emission from a compact jet, a scenario further supported by a number of evidences: the broken power-law shape of the continuum distribution in the radio-to-UV range; the $\sim 4\%$ degree of polarisation measured in the nuclear mid-IR continuum, together with the mild optical extinction ($A_V \sim 1\, \rm{mag}$); and the ``harder when brighter'' behaviour of the X-ray spectrum, indicative of self-Compton synchrotron radiation. A remarkable feature is the steepness of the IR-to-UV core continuum, characterised by a power-law index of $\sim 2.6$, as compared to the canonical value of $0.7$. Alternatively, to explain the interferometric data by thermal emission would require an exceptionally compact dust distribution when compared to those observed in nearby AGN, with $A_V \gtrsim 2.8\, \rm{mag}$ to account for the IR polarisation. This is in contrast with several observational evidences against a high extinction along the line of sight, including the detection of the nucleus in the UV range and the well defined shape of the power-law continuum. The case of NGC\,1052 shows that compact jets can dominate the nuclear emission in LLAGN across the whole electromagnetic spectrum, a scenario that might be common among this class of active nuclei.
\end{abstract}




\section{Introduction}\label{intro}

Low-Luminosity Active Galactic Nuclei (LLAGN) are the most common population of active nuclei, including about one third of all galaxies in the local Universe \citep{1997ApJS..112..315H,2008ARA&A..46..475H}. In spite of their name, LLAGN are not simply scaled-down counterparts ($L_{\rm bol} \lesssim 10^{42}\, \rm{erg\,s^{-1}}$) of the brighter Seyfert nuclei and quasars, but above all they are radiatively inefficient nuclei with typical $\log(L_{\rm bol}/L_{\rm edd}) \lesssim - 3$. Their low sub-Eddington luminosities have been ascribed to a major change in the inner accretion flow, when the low accretion rate originates a Radiatively Inefficient Accretion Flow (RIAF) below a certain truncation radius in the disc \citep{1995Natur.374..623N,1995MNRAS.277L..55F,2003ApJ...582..133D}. Further structural changes are expected at low luminosities, since the wind driven by a weak nucleus would not be able to sustain either the broad-line clouds \citep{2003ApJ...589L..13N,2009ApJ...701L..91E} or the putative AGN torus \citep{2006ApJ...648L.101E,2007MNRAS.380.1172H}. The latter is the cornerstone of the Unified Model to explain the dichotomy between type 1 and type 2 Seyfert nuclei \citep{1993ARA&A..31..473A,1995PASP..107..803U}, but its possible disappearance at low luminosities opens one of the most compelling questions in the field: what is the nature of the continuum emission in LLAGN?

This has been a source of debate in the recent years, partly due to the heterogeneous nature of the LLAGN class. At radio and X-ray wavelengths, where the host galaxy contribution is negligible, the dominant nuclear emission was early associated with non-thermal processes \citep{1978ApJ...221..456C,1978Natur.276..374O,1980A&A....87..152H,2001ApJ...549L..51H,2003ApJ...583..145T}. However, the optical/UV continuum and the ionisation mechanisms in LLAGN stimulated an intense discussion over the last $\sim 40$ years \citep{1976ApJ...203L..49K,1983ApJ...266L..89K,1985ApJ...289..475F,1985MNRAS.213..841T,2000ApJ...529..219S,2007MNRAS.377.1696M}. In the mid-IR range, the detection of silicate emission within the inner few tens of parsecs (e.g. \citealt{2010A&A...515A..23H}, \citealt{2012AJ....144...11M}, \citealt{2013ApJ...777..164M}) and the mid-IR\,--\,X-ray ratios consistent with those in Seyfert galaxies \citep{2015MNRAS.454..766A} point to the presence of circumnuclear dust in these nuclei. However, the power-law continuum found in a number of LLAGN suggests that non-thermal radiation may represent a major contribution to the energy output (\citealt{1982ApJ...252L..53R}, \citealt{1996ApJ...462..183H}, \citealt{1999ApJ...516..672H}, \citealt{2012AJ....144...11M}, \citealt{2013ApJ...777..164M}, \mbox{\citealt{2012JPhCS.372a2006F}}), associated with a RIAF (e.g. \mbox{\citealt{2014MNRAS.438.2804N}}) or a compact jet (M81 in \citealt{2008ApJ...681..905M}; M87 in \citealt{2016MNRAS.457.3801P}).

Interferometry is the only technique that allows to probe the spatial distribution of the mid-IR emission at sub-parsec scales in nearby AGN \citep[e.g.][]{2009A&A...502...67T,2011A&A...531A..99T,2013A&A...558A.149B,2014A&A...565A..71L}, and investigate the torus endurance at low luminosities. However, only a few LLAGN are bright enough to reach the instrumental limit of the MID-Infrared Interferometric Instrument (MIDI, correlated flux $\gtrsim 90\, \rm{mJy}$; \citealt{2013A&A...558A.149B}), located at the Very Large Telescope Interferometer (VLTI) on Cerro Paranal (Chile). This is the most sensitive interferometric instrument in the mid-IR to date. One of these sources is NGC\,1052, a nearby E4 galaxy ($z = 0.0045$; $D = 18\, \rm{Mpc}$, $1\,\arcsec \sim 87\,\mbox{pc}$, \citealt{2003ApJ...583..712J}) that hosts a prototypical Low-Ionisation Nuclear Emission-line Region (LINER) and has been considered a reference to define the general properties of its class \citep{1976ApJ...203L..49K,1978MNRAS.183..549F,1980A&A....87..152H}. With a bolometric luminosity of $L_{\rm bol} \sim 7 \times 10^{42}\, \rm{erg\,s^{-1}}$ and $\log(L_{\rm bol}/L_{\rm edd}) = -3.4$ (\citealt{2012JPhCS.372a2006F}, \citealt{2018MNRAS.478L.122R}; see Section\,\ref{sed} in this work), NGC\,1052 is close to the domain of Seyfert nuclei, possibly in a transition zone between the two classes.

Observational evidences point to the presence of a nuclear absorber in this source: the detection of broad H$\alpha$ emission lines in polarised light from the central region \citep{1999ApJ...525..673B}, the high hydrogen absorption column derived from X-ray observations ($N_{\rm H} \sim 10^{23}\, \rm{cm^{-2}}$; \citealt{2014A&A...569A..26H}), and the detection of HCN molecular absorption \citep{2016ApJ...830L...3S}. However, the opposite is suggested by the presence of broad components in H$\alpha$ and H$\beta$ also in the intensity (unpolarised) spectrum \citep{1999ApJ...525..673B,2005ApJ...629..131S}, the detection of the $9.7\, \rm{\micron}$ silicate band in emission \citep{2008ApJ...684..270K}, typical of type 1 AGN, the mild $A_V = 1.05\, \rm{mag}$ attenuation measured within the inner $0\farcs8$ in an \textit{HST}/FOS spectrum \citep{2015ApJ...801...42D}, and the variability of the nuclear UV continuum between \textit{HST} imaging and spectra at different epochs \citep{2005ApJ...625..699M}, which makes unlikely the presence of a deeply obscured nucleus in this galaxy. On the other hand, one of the most remarkable features of NGC\,1052 is the shape of the nuclear continuum distribution, following a broken power law in the IR-to-UV range \citep{1982ApJ...252L..53R,1982ApJ...263..624B,2012JPhCS.372a2006F,2015ApJ...814..139K}. This is indicative of self-absorbed synchrotron emission from a compact jet \citep{1979ApJ...232...34B}, which could dominate the overall energy output in this nucleus.

The aim of this work is to probe the mid-IR brightness distribution in the nucleus of NGC\,1052 using VLTI/MIDI observations and discriminate between the proposed scenarios for the origin of this emission. That is, a nuclear IR source resolved by MIDI would imply the survival of the torus at low luminosities, while a non-thermal source is expected to be unresolved. The latter would originate in a very small region where the particles are accelerated, e.g. close to the base of the jet, the corona, or the shock location \citep{1979ApJ...232...34B,2005ApJ...635.1203M,2008Natur.452..966M,2014MNRAS.438..959P}. To discern the thermal or non-thermal nature of the emission, we acquired also \textit{Ks}- and \textit{N}-band imaging in polarised light using the Long-slit Intermediate Resolution Infrared Spectrograph (LIRIS) at the William Herschel Telescope (WHT, $4.2\, \rm{m}$) and CanariCam \citep{Telesco:2003aa} at the Gran Telescopio Canarias (GTC, 10.4\, \rm{m}), both facilities located at the Roque de los Muchachos Observatory (La Palma, Spain).

This work is organised as follows. In Section\,\ref{data} we present the spectral flux distribution, the mid-IR interferometric observations, and the IR polarimetry for the nucleus of NGC\,1052. In Section\,\ref{results} we analyse the brightness distribution of the mid-IR emission in the nucleus of NGC\,1052. In Section\,\ref{discuss} we compare the case of NGC\,1052 with the Circinus Galaxy --\,the source with the best ($u,v$) plane coverage using IR interferometric observations\,-- and discuss the nature of the continuum emission on the basis of its nuclear flux distribution. The final summary and conclusions are presented in Section\,\ref{sum}.

\begin{figure}
   \centering
   \includegraphics[width=\columnwidth]{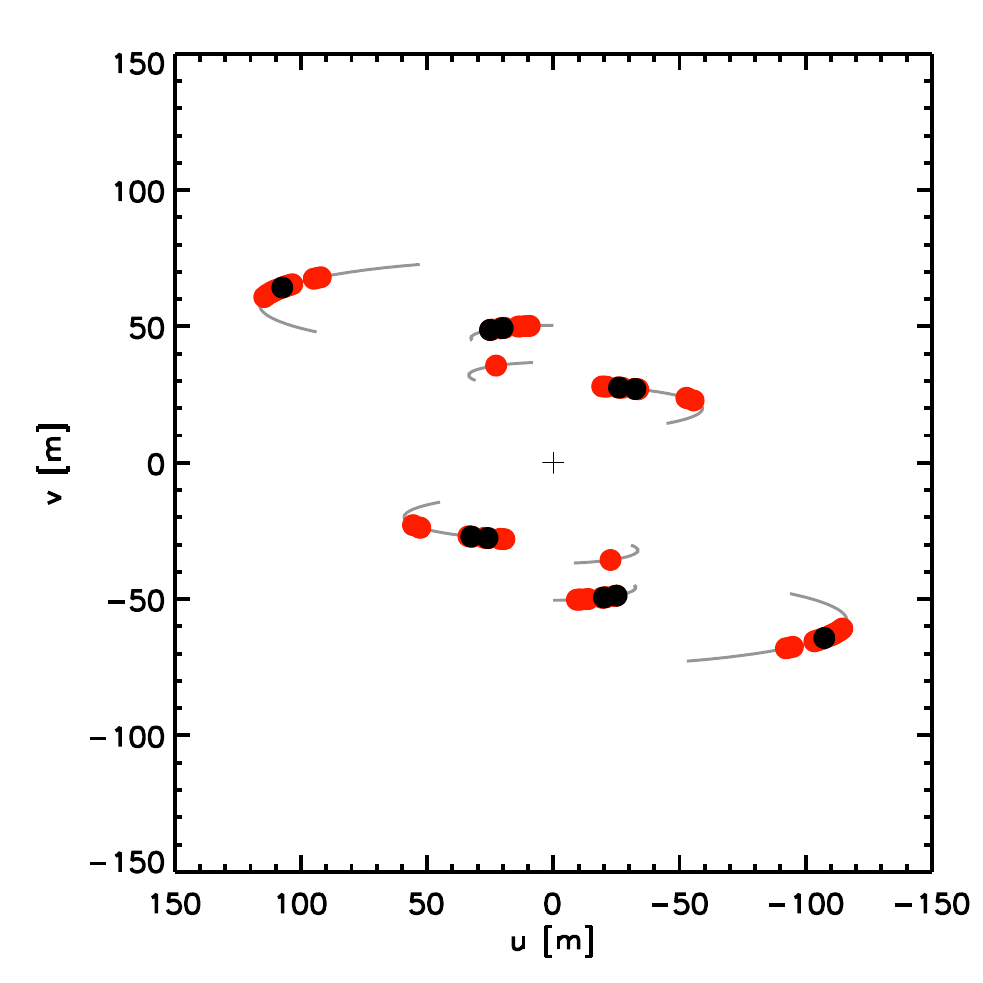}
   \caption{($u, v$) coverage for NGC\,1052 for the VLTI/MIDI observations. Five independent and well calibrated interferometric measurements were obtained (black dots). Data acquired at too large OPD drift values ($\gtrsim 100\, \rm{\micron}$) or missing the ozone atmospheric absorption feature at $9.6\, \rm{\micron}$ are flagged in red colour.}\label{fig_uvplane}
\end{figure}

\section{Dataset}\label{data}

Four different datasets are used in the present study. The first one consists of mid-IR interferometric observations acquired with VLTI/MIDI (Section\,\ref{vlti}), which probe the brightness distribution at the highest angular resolution available so far. The second dataset includes \textit{Spitzer}/IRS and VLT/VISIR\footnote{V\textsc{lt} Imager and Spectrometer for mid-InfraRed \citep{2004Msngr.117...12L}.} spectra to investigate the dust features in the mid-IR range (Section\,\ref{mirspec}). The third dataset adds broad- and intermediate-band IR polarimetric imaging in the \textit{K} and \textit{N}~bands. The fourth dataset is a high-angular resolution spectral flux distribution assembled using subarcsec resolution flux measurements ($\sim 15$--$50\, \rm{pc}$), compared with a second flux distribution based on low-angular resolution measurements to characterise the host galaxy contamination (Section\,\ref{sed}). The flux tables for both distributions are provided in Appendix\,\ref{app_sed}.

\subsection{Mid-IR interferometry}\label{vlti}
Interferometric observations of NGC\,1052 were obtained with VLTI/MIDI, a Michelson interferometer that combines the light beams from two 8.2\,m Unit Telescopes (UTs) in the \textit{N}~band ($8$\,--\,$13\, \rm{\micron}$; \citealt{2003Ap&SS.286...73L}). The instrument delivers two main interferometric observables from the interference pattern: the correlated flux spectra and the differential phases. Visibilities are derived as the ratio of the correlated flux over the total or photometric flux. However, the use of correlated fluxes is more convenient for faint sources (< 200\,mJy), since measuring photometric fluxes against the fluctuations of the bright sky can be particularly challenging for these targets \mbox{\citep{2012SPIE.8445E..1GB}}. Therefore in our analysis we use the measured correlated fluxes instead of the visibilities.

The MIDI interferometric campaign of NGC\,1052 was carried out during the nights of August 6th and 7th, and November 2nd, 3rd, and 4th, 2014 (Programme ID: 093.B-0616, 094.B-0918). Our observing strategy, data reduction process, and analysis of the data were planned and performed using previous experience acquired during the MIDI AGN Large Programme \citep{2012SPIE.8445E..1GB}. The low-spectral resolution NaCl prism ($R\equiv\lambda/\Delta \lambda \sim 30$) was used to disperse the light of the beams. A complete log of the observations and the instrumental setup can be found in Appendix\,\ref{app_log}.

With a total flux of $\sim 130\, \rm{mJy}$ at $12\, \rm{\micron}$ within an aperture radius of $0\farcs4$ (VLT/VISIR, see Table\,\ref{sed_high}), the nucleus of NGC\,1052 is well below the sensitivity limit for the standard modes of MIDI. At such low fluxes, the fringe tracker can no longer follow the interferometric signal because strong noise peaks are too frequent for the group-delay tracking to stay on the MIDI fringe. Therefore we used the \textit{no-track} mode, i.e. switching off the group-delay tracking to scan the fringes blindly. This is the only observing mode that allows the detection of faint targets below $\sim 100\, \rm{mJy}$ in correlated flux. For each scan, the fringes were recorded close to the optical path delay (OPD) measured for the calibrators (HD\,10380 and HD\,18322), which were observed recurrently every $30$--$50\, \rm{min.}$ in order to correct for possible OPD drifts. The widest baselines ($\sim 130\, \rm{m}$) were included in order to test the finest angular scales feasible for the UTs.

MIDI observations were processed and reduced using the \textit{MIDI Interactive Analysis and Expert Work Station} software \citep[\textsc{mia+ews}\footnote{\textsc{ews} is available for download from: \hfill \\ \url{http://home.strw.leidenuniv.nl/~jaffe/ews/index.html}},][]{2004SPIE.5491..715J}, which implements the coherent integration method for MIDI data. For each measurement we derived the ratio of the target signal over the calibrator signal, which was then multiplied by the know flux of the calibrator to obtain the correlated fluxes. The spectral templates of \citet{1999AJ....117.1864C} were used for the calibrators HD\,10380 and HD\,18322. However, further processing is needed to correct for the correlation losses, a known bias that can be particularly severe for faint targets and cannot be corrected using the standard calibration techniques. Correlation losses are caused by the larger dispersion in the group delay estimates of weak sources when compared to the bright ones, and results in the underestimation of the correlated fluxes derived. The flux loss can be significant ($> 10\%$) for sources below $200\, \rm{mJy}$, especially at the shortest wavelengths where the group delay error propagates into a larger phase error. To correct for correlation losses caused by the scatter in the atmospheric phase we performed \textit{dilution} experiments as in the MIDI AGN Large Programme (\citealt{2012SPIE.8445E..1GB}, \citealt{2013A&A...558A.149B}; see also \citealt{2018A&A...611A..46F} for the case of IC\,3639). The final $(u, v)$ coverage is shown in Fig.\,\ref{fig_uvplane}. We obtained a total of 5 independent and well calibrated interferometric measurements (black dots). Observations obtained within 1 hour apart and sharing the same calibrator were added together using the same stacking method as that described in \citet{2013A&A...558A.149B}. Data acquired at too large OPD drift values ($\gtrsim 100\, \rm{\micron}$) are flagged in red colour in Fig.\,\ref{fig_uvplane}. Additionally, we removed those spectra where the ozone atmospheric absorption feature at $9.6 \, \rm{\micron}$ was not detected, following the method described in \citet{2012SPIE.8445E..1GB}. Due to the low correlated fluxes ($\gtrsim 100\, \rm{mJy}$) and thus the low signal-to-noise ratio ($S/N$), the full spectral array could not be recovered.

\subsection{Mid-IR spectroscopy}\label{mirspec}
\begin{figure}
   \centering
   \includegraphics[width = \columnwidth]{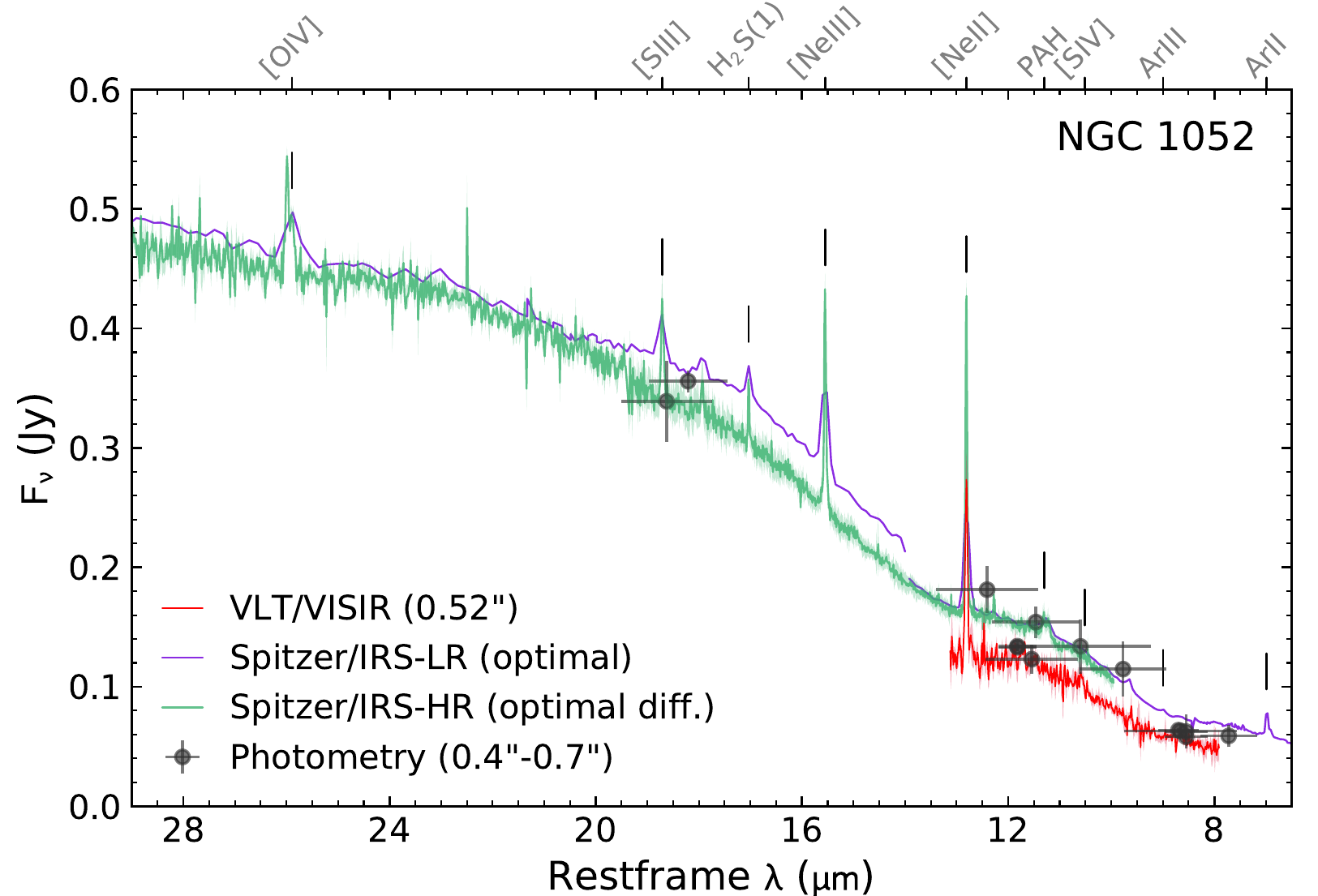}
   \caption{Nuclear IR spectra of NGC\,1052 compiled from previous VISIR observations (in red) and the CASSIS database. The latter includes \textit{Spitzer}/IRS spectra in the high-spectral resolution mode (in green; ``optimal-differential'' extraction) and in the low-spectral resolution mode (in purple; ``optimal'' extraction). The discontinuity in the low-resolution spectrum at $\sim 14\, \rm{\micron}$ is possibly caused by errors in the calibration between the short-low and the long-low modules. For comparison, photometric measurements at subarcsec resolution ($0\farcs4$--$0\farcs7$) are also shown as black dots. The main emission lines and PAH bands are marked in the spectrum.}\label{fig_irspec}
\end{figure}

Due to a refurbishment of VISIR during the interferometric campaign of NGC\,1052, a simultaneous mid-IR spectrum with this instrument to double-check the single-dish total flux calibration could not be acquired. However previous VISIR spectroscopy was available in the ESO scientific archive, observed on the nights of October 21st, November 8th, and December 23rd and 24th, 2010 (PI: D. Asmus; Programme ID: 086.B-0349). These were taken in the low-resolution mode ($R \sim 350$) using the four spectral settings available prior to the VISIR refurbishment ($8.5$, $9.8$, $11.4$, and $12.4\, \rm{\micron}$). The ESO Common Pipeline Library (\textsc{CPL} v5.3.1, VISIR recipe v3.4.4) was used to reduce and calibrate the data (calibrators: HD\,25025, HD\,4128, HD\,23249, and HD\,8512), with further post-processing following the method described in \citet{2010A&A...515A..23H}. Finally, the spectrum was extracted using a $0\farcs75 \times 0\farcs52$ slit, i.e. identical to the aperture used by MIDI. The spectrum is shown in red colour in Figs\,\ref{fig_irspec} and \ref{fig_sed}.

Additionally, mid-IR spectra at high- and low-spectral resolution were collected from the Cornell Atlas of \textit{Spitzer}/Infrared Spectrograph Sources (CASSIS; \citealt{2015ApJS..218...21L}). The $R \sim 600$ spectrum (PI: C. Leitherer; AOR key: 11510784) covers the $9.9$--$37.1\, \rm{\micron}$ range with slit sizes of $4\farcs7 \times 11\farcs3$ and $11\farcs1 \times 22\farcs3$ in the SH ($9.9$--$19.5\, \rm{\micron}$) and LH modes ($18.8$--$37.1\, \rm{\micron}$), respectively. Fig.\,\ref{fig_irspec} shows the ``optimal differential'' spectrum extracted using the point-spread function profile (PSF; dark green). This method is optimal to recover the spectra of unresolved sources mixed with extended emission. The $R = 60$--$130$ spectrum (PI: H. Kaneda; AOR key: 18258688), shown in purple colour, corresponds to the ``optimal'' extraction and covers the $5.2$--$14.5\, \rm{\micron}$ and $14.0$--$38.0\, \rm{\micron}$ ranges with slit sizes of $3\farcs7 \times 57''$ and $10\farcs7 \times 168''$, respectively.

The infrared spectra are compared in Fig.\,\ref{fig_irspec} with subarcsec photometry at $\sim 0\farcs5$ resolution. Most of these fluxes correspond to VLT/VISIR and Gemini/T-ReCS\footnote{Thermal-Region Camera Spectrograph.} observations published in the literature (see Section\,\ref{sed} for a detailed description; Table\,\ref{sed_high}). Overall, the continuum level in the three infrared spectra is in fair agreement with the photometric data, despite the large difference in the aperture sizes, up to a factor $\sim 20$ between \textit{Spitzer}/IRS and VISIR at $18\, \rm{\micron}$. A scatter in the mid-IR photometry of the order of $\sim 15\%$ is seen in the $10$--$13\, \rm{\micron}$ range, possibly caused by the contribution of the polycyclic aromatic hydrocarbon (PAH) feature at $11.3\, \rm{\micron}$ and the fine-structure lines of [\ion{Ne}{ii}]$_{\rm 12.8 \micron}$ and [\ion{S}{iv}]$_{\rm 10.5 \micron}$ to the VISIR filters B11.7, B12.4, and B10.7, respectively. This is confirmed by the comparison of flux measurements in pure continuum emission filters --\,e.g. the PAH2\_2 filter in VISIR for two different epochs and the Si5 filter in CanariCam, which are fully consistent within the uncertainties. The $8$--$13\, \rm{\micron}$ continuum in the \textit{Spitzer}/IRS spectra is $20$--$40\, \rm{mJy}$ higher when compared to the VISIR spectrum ($\sim 15$--$25\%$). The silicate band at $9.7\, \rm{\micron}$ is detected by both VISIR and \textit{Spitzer} and peaks in the $10.5$--$11\, \rm{\micron}$ range, a shift typically seen when the feature is in emission \citep{2015ApJ...803..110H}. No significant difference is detected in the strength of the silicate feature between both spectra ($S_{\rm sil} = \ln(F_{\rm \lambda_{peak}} / F_{\rm \lambda_{cont}}) \sim 0.15$). The $18\, \rm{\micron}$ silicate feature is detected in emission in the \textit{Spitzer}/IRS-LR spectrum, which shows higher fluxes in this range than the IRS-HR spectrum.

\subsection{IR polarimetry}\label{pol}
To investigate further the nature of the IR emission in NGC\,1052, we took advantage of the polarimetric capabilities of WHT/LIRIS and GTC/CanariCam. NGC\,1052 was observed on the nights of January 14th, 2014 (Programme ID: WHT23-13B) and September 16th, 2013 (Programme ID: GTC30-13B), using the \textit{Ks}~band (cut-on at $1.99\, \rm{\micron}$, cut-off at $2.31\, \rm{\micron}$) and the \textit{N}~band intermediate-band filters Si2 ($\lambda_{\mbox{\tiny c}} = 8.7\, \rm{\micron}$, $\Delta \lambda = 1.1\, \rm{\micron}$ at $50\%$ cut-on/off) and Si5 ($\lambda_{\mbox{\tiny c}} = 11.6\, \rm{\micron}$, $\Delta \lambda = 0.9\, \rm{\micron}$ at $50\%$ cut-on/off), respectively.

Near-IR images were acquired following a dithering plus offset pattern, where the location of NGC\,1052 on the detector was alternated between the centre and the sides of the detector field-of-view. Using a Wedged Double Wollaston system (WeDoWo; \citealt{1997A&AS..123..589O}), LIRIS splits the light beam in four slices, delivering simultaneously four different position angles ($0$, $90$, $45$, and $135\, \rm{deg}$) for the polarised flux. Each acquisition had an exposure time of 25\,s, and were acquired following a three-point dither pattern to improve the subtraction of the background. The total exposure time on source was $1200$\,s, that is $600$\,s for each of the two orthogonal positions of the half-wave plate, resulting in eight final images containing the information on the I, U, and Q parameters. The rotation of the retarder plate allows us to switch the positions of the ordinary and extraordinary beams on the detector, and thus cancel possible differences in the transmission along the light path. To calibrate and correct for the instrumental polarisation, two zero-polarisation standard stars (HD\,14069, GD\,319; \citet{1992AJ....104.1563S}) and a linearly polarised star (HD\,3856C; \citealt{1992ApJ...386..562W}) were observed during the same night. The images were reduced using the dedicated \texttt{lirisdr}\footnote{\url{https://github.com/jaacostap/lirisdr}} package in \textsc{iraf}\footnote{\textsc{iraf} is distributed by the National Optical Astronomy Observatories, which are operated by the Association of Universities for Research in Astronomy, Inc., under cooperative agreement with the National Science Foundation.}, and includes the background subtraction, alignment and combination of the individual frames. Nuclear fluxes for each frame were derived using aperture photometry ($r \sim 0\farcs75$), subtracting the adjacent background dominated by the underlying galaxy. The Stokes parameters (I, Q, and U) were derived from the measured fluxes following the method described in \citet{2011AJ....142...33A}. After correcting for the polarisation bias we measured a linear polarisation degree in the nucleus of $\sim 0.45 \pm 0.07 \%$ at PA\,$= 169\degr \pm 9\degr$ in the \textit{K}~band. For the polarised standard we obtained $2.2 \pm 0.2\%$ and $75\degr \pm 4\degr$, in very good agreement with the $2.21 \pm 0.55 \%$ and $78\degr \pm 17\degr$ values given by \citet{1992ApJ...386..562W}.

In the mid-IR, the background subtraction was performed in real time during the observations using the standard chop-nod technique, with a chop-throw of $9''$, a chop-angle oriented along the North-South axis, and a chop-frequency of $1.93\, \rm{Hz}$. The instrumental position angle (IPA), i.e. the orientation of the detector vertical axis on the sky, was $0\degr$ East of North, and the telescope was nodded every $46$\,s along the chopping direction. Only one ``slot'' was used during the observations, i.e. one slice of the CanariCam polarimetry mask ($25\farcs6 \times 2\farcs0$). For each filter, four observing cycles were completed with a total on-source time of $1748$\,s and $1776$\,s at $8.7\, \rm{\micron}$ and $11.6\, \rm{\micron}$, respectively. The data were reduced following the method described in \citet{2016MNRAS.458.3851L,2018MNRAS.478.2350L}, which includes the correction of the I, Q, and U Stokes parameters by the instrumental polarisation, the polarisation efficiency, and the polarisation bias. Observations of the unpolarised standard star HD\,16212 were used to calibrate: \textit{i)} the instrumental linear polarisation, of about $0.6 \pm 0.1 \%$ and $0.5 \pm 0.1 \%$ at $8.7\, \rm{\micron}$ and $11.6\, \rm{\micron}$, respectively; \textit{ii)} the shape of the PSF, characterised by a full-width at half-maximum (FWHM) of $0\farcs45 \times 0\farcs38$ with PA\,$= 50\degr$ at $8.7\, \rm{\micron}$ and $0\farcs36 \times 0\farcs34$ with PA\,$ = 38\degr$ at $11.6\, \rm{\micron}$; and \textit{iii)} the absolute flux of the mid-IR core in NGC\,1052, using the catalogue fluxes given by \citet{1999AJ....117.1864C}. Typical flux calibration uncertainties of $15\%$ were estimated.

Due to the limited number of mid-IR polarimetric standards for a 10\,m class telescope, we used as calibrator the Herbig-Haro object HH7-11 ($\alpha =$\,03h\,29m\,3.75s, $\delta =$\,+31d\,16m\,4.0s), which shows a \textit{K}~band linear polarisation degree of $7.24\%$ at a position angle PA\,$= 56\degr \pm 4\degr$ \citep{1976AJ.....81..314S,1988MNRAS.231..445T}. Thus, to calibrate the PA zero-offset for the linear polarisation, HH7-11 was observed immediately before NGC\,1052, with an on-source time of $146$\,s. To our knowledge, these are the first mid-IR polarimetric observations of HH7-11 published in the literature. Once corrected by the instrumental polarisation and the polarisation efficiency, we measure a degree of $2.8 \pm 0.4 \%$ and $6.5 \pm 0.5 \%$ at $8.7\, \rm{\micron}$ and $11.6\, \rm{\micron}$, respectively. The uncertainties in the degree and the PA of the polarisation were estimated following the method of \mbox{\citet{1993A&A...274..968N}}, taking into account the signal-to-noise, the measured degree, and the PA values. To estimate the zero-angle, we assumed that the angle of the linear polarisation in HH7-11 is constant from $2.2\, \rm{\micron}$ to $11.6\, \rm{\micron}$, which suggests that the same mechanism dominates the polarised emission in this wavelength range. Finally, the zero-angle of the linear polarisation was obtained from the difference between the $\theta_{\mbox{\tiny 2.2 \micron}} = 56\degr$ value and our measurements of $\theta_{\mbox{\tiny 8.7 \micron}} = 38\degr \pm 6\degr$ and $\theta_{\mbox{\tiny 11.6 \micron}} = 40\degr \pm 3\degr$. That is, we obtain zero-offset corrections to the PA of the linear polarisation of about $\Delta\theta_{\mbox{\tiny 8.7 \micron}} = 18\degr \pm 6\degr$ and $\Delta\theta_{\mbox{\tiny 11.6 \micron}} = 16\degr \pm 4\degr$.

To minimise a possible contamination from extended (diffuse) warm dust emission, we measured the total and polarised fluxes for the nucleus of NGC\,1052 within a $0\farcs5$ ($42.5\, \rm{pc}$) aperture radius. The polarisation degree and position angle measured are $4.4 \pm 0.5\%$ at $125\degr \pm 4\degr$ in the Si2 filter, and $3.5 \pm 0.6\%$ at $118\degr \pm 5\degr$ for Si5.

\begin{figure*}
   \centering
   \includegraphics[width=\textwidth]{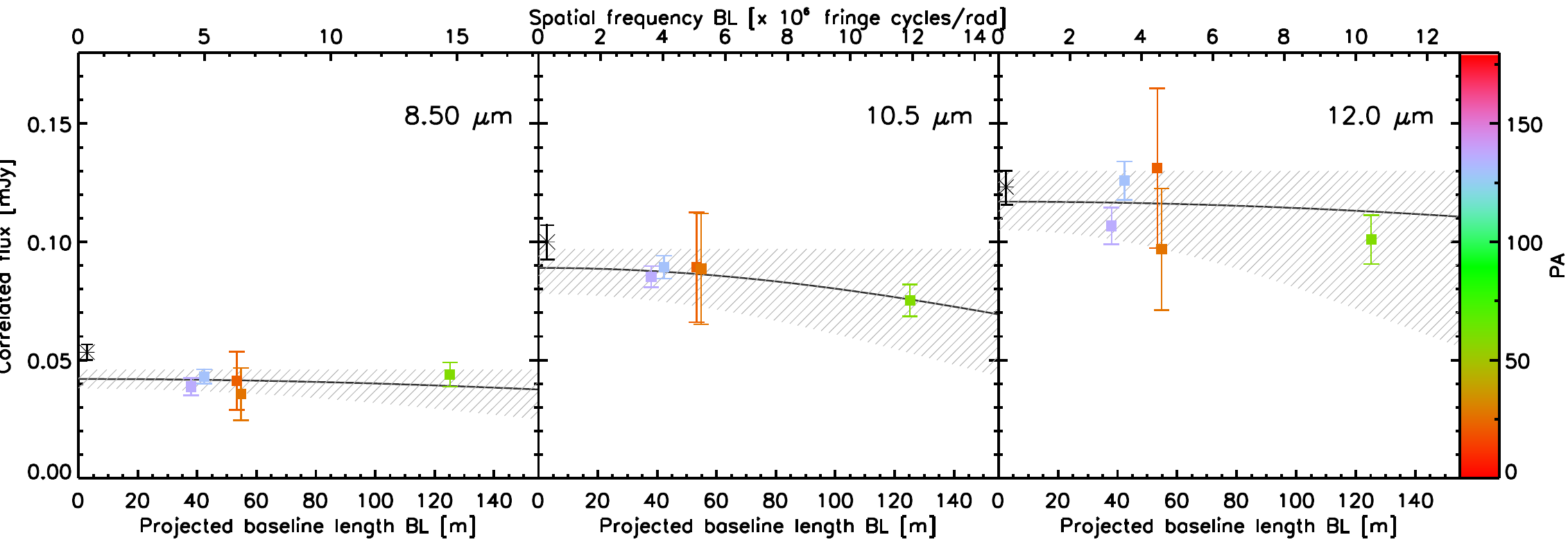}
   \caption{VLTI/MIDI correlated fluxes for NGC\,1052 as a function of the projected baseline length \textit{BL} at $\lambda = 8.5\, \rm{\micron}$ (left panel), $10.5\, \rm{\micron}$ (centre), and $12.0\, \rm{\micron}$ (right). Data points are colour coded based on the position angle of the corresponding projected baseline. Black asterisks at zero-metre baseline length represent the single-dish total flux measurements (see Table\,\ref{tab_midi}). The solid line represents the best-fit Gaussian model and the shaded region corresponds to its 1$\sigma$ uncertainty area. The correlated fluxes at $8.5\, \rm{\micron}$ show the presence of an unresolved source ($< 3.9\, \rm{mas} \sim 0.3\, \rm{pc}$) in the nucleus of NGC\,1052, while at $10.5\, \rm{\micron}$ and $12.0\, \rm{\micron}$ the nucleus could be marginally resolved.
}\label{fig_radial}
\end{figure*}

\subsection{Spectral Flux Distribution}\label{sed}
Flux measurements based on high-angular resolution observations are used to isolate the nuclear emission in NGC\,1052 and avoid possible contamination by the host galaxy. This dataset includes $0\farcs2$ aperture radius photometry from near-IR adaptive optics-assisted observations with VLT/NaCo\footnote{Nasmyth Adaptive Optics System (NAOS) + Near-Infrared Imager and Spectrograph \citep[CONICA;][]{2003SPIE.4841..944L,2003SPIE.4839..140R}.} and diffraction-limited imaging with \textit{HST} in the UV/optical range. In all cases, the local background was estimated in a surrounding annulus and subtracted from the measured fluxes. The flux distribution of NGC\,1052 presented here includes a few updates with regard to previous versions introduced in \citet{2012JPhCS.372a2006F}, \citet{2015ApJ...814..139K}, and more recently in \citet{2018MNRAS.478L.122R}. Specifically, the mid-IR range has been completed with VLT/VISIR (observations made between September 2009 and October 2010) and Gemini/T-ReCS (acquired in August 2007) flux measurements at $\sim 0\farcs5$ resolution from \citet{2014MNRAS.439.1648A}. In this work we also provide previously unpublished VLT/VISIR photometry with the PAH1 and PAH2\_2 filters (Programme ID: 095.B-0376; PI: D. Asmus) for observations obtained with the upgraded VISIR instrument \citep{2015Msngr.159...15K,2016SPIE.9908E..0DK} using exposure times of $7$ and $10$\,min., respectively. These include three different acquisitions between July and August 2015, and were intended to check for possible flux changes since the MIDI data were acquired. The reduction was performed with the standard ESO pipeline, and the fluxes were extracted using MIRphot \citep[see][]{2014MNRAS.439.1648A}. The resulting fluxes for the PAH1 ($58 \pm 3\, \rm{mJy}$) and PAH2\_2 filters ($133 \pm 3\, \rm{mJy}$) are in agreement with previous measurements. Thus, no variability at subarcsec scales in the mid-IR range is detected between $2007$ and $2015$. Additionally, the unresolved nuclear source in GTC/CanariCam total intensity images ($< 0\farcs43 \sim 38\, \rm{pc}$) has a flux of $63 \pm 6\, \rm{mJy}$ and $123 \pm 12\, \rm{mJy}$ in the Si2 and Si5 filters, respectively, in agreement with previous VISIR and T-ReCS measurements (see Table\,\ref{sed_high}).

On the other hand, NGC\,1052 is much more active in the millimetre range showing a high variability ($\sim 70\%$) at $22$ and $43\, \rm{GHz}$ during a VLBI monitoring campaign \citep{2019arXiv190102639B}. The maximum and minimum fluxes measured between 2005 and 2009 have been incorporated to our high-angular resolution database. At higher frequencies we also included Atacama Large Millimeter/submillimeter Array (ALMA) measurements in the $108$--$353\, \rm{GHz}$ range recently published by \citet{2019arXiv190102280P}, and archival observations in the $419$--$478\, \rm{GHz}$ range where NGC\,1052 was acquired as a calibrator (programme: 2013.1.00749.S, PI: R. Dominik). We calibrated the archival observations using \textsc{casa}\footnote{Common Astronomy Software Applications package} v4.7.2 and the scripts provided by the ALMA Observatory in the scientific archive. Four continuum bands were reconstructed using \textsc{casa} v5.4.0, with \texttt{briggs} weighting and a robust parameter of $0.5$ for the cleaning process. The synthesised beam has a size of $\sim 0\farcs12$ (see Table\,\ref{sed_high}), thus comparable with the optical and near-IR data. The nuclear flux was measured by fitting a 2D elliptical Gaussian profile to the nuclear source, which remain unresolved in all bands.

IR-to-UV fluxes were corrected for Galactic reddening ($A_{V} = 0.073\, \rm{mag}$; \citealt{2011ApJ...737..103S}), using the extinction curve from \citet{1989ApJ...345..245C} and $R_V = 3.1$. Fluxes derived from X-ray and radio interferometric observations were compiled from published values in the literature. The complete high-angular resolution flux distribution\footnote{Tabulated fluxes are not corrected for Galactic reddening.} and the corresponding references are given in Table\,\ref{sed_high}. Although X-ray observations do not have subarcsec resolution, a significant contamination from the underlying galaxy is not expected in this range, thus we assume that X-ray fluxes are representative of the LLAGN core. Low-angular resolution measurements ($\gtrsim 1''$; Table\,\ref{sed_low}) were compiled from the NED database\footnote{The NASA/IPAC Extragalactic Database (NED) is operated by the Jet Propulsion Laboratory, California Institute of Technology, under contract with the National Aeronautics and Space Administration.} and the literature, and are used as a reference for the contribution of the host galaxy. A numerical integration of the high-angular resolution flux distribution plus low-angular resolution measurements in the $20\, \rm{\micron}$--$3\, \rm{mm}$ gap gives a luminosity of $5.3\, \times 10^{42}\, \rm{erg\,s^{-1}}$ for the radio-to-UV range \citep{2018MNRAS.478L.122R}, and $1.7 \times 10^{42}\, \rm{erg\,s^{-1}}$ for the X-ray range. Thus, we adopt $L_{\rm bol} \sim 7 \times 10^{42}\, \rm{erg\,s^{-1}}$, which corresponds to $\log(L_{\rm bol}/L_{\rm edd}) = -3.4$ for a black hole mass of $1.55 \times 10^{8}\, \rm{M_\odot}$ \citep{2002ApJ...579..530W}.

\section{Results}\label{results}
Despite the lack of the full spectral information (Section\,\ref{vlti}), a hint on the \textit{N}~band shape can be recovered from the average correlated fluxes at $8.5$, $10.5$, and $12\, \rm{\micron}$, derived for every independent ($u,v$)-point. Prior to the modelling in Sections\,\ref{gaussian} and \ref{plaw}, one can anticipate from Fig.\,\ref{fig_radial} that the correlated flux does not vary significantly with the projected baseline length. A somewhat lower correlated flux at the widest baselines ($130\, \rm{m}$) might be present at $12\, \rm{\micron}$ --\,perhaps also at $10.5\, \rm{\micron}$\,-- but the difference is smaller than the error bars. This implies that \textit{the mid-IR brightness distribution in the nucleus of NGC\,1052 is compact at the highest resolution provided by MIDI using the baselines available for the UTs}. This compact mid-IR core dominates the total emission of the galaxy at the larger scales seen by \textit{Spitzer} ($> 10''$) and remains unresolved even at an angular resolution about $2000$ times higher.

\subsection{Gaussian brightness distribution}\label{gaussian}
Fitting the correlated fluxes with a Gaussian distribution is a common strategy to recover, at first order, the size of the mid-IR brightness distribution \citep{2013A&A...558A.149B}. Due to the limited number of interferometric measurements in NGC\,1052, a simple model was used in the first analysis to avoid possible degeneracies. 2D symmetrical Gaussian distributions were fitted to each of the three wavelength bands, using two free parameters: the total flux ($F_\lambda$) and the FWHM ($\theta_\lambda$). Single-dish fluxes, mainly from VISIR, were included to sample the zero-metre baseline length and constrain the total integrated flux of the nucleus (``VISIR spec.'' column in Table\,\ref{tab_midi} and black asterisks in Fig.\,\ref{fig_radial}).

\begin{table}
  \caption{Best-fit parameters obtained for a Gaussian model applied to the measured visibilities for the three wavelength bands extracted from MIDI spectra. The columns correspond to the FWHM of the Gaussian (given in $\rm{mas}$ and $\rm{pc}$, for $D = 18\, \rm{Mpc}$), the total integrated flux of the model, the best-fit reduced $\chi^2$, and the average continuum flux in the VISIR spectrum (the error corresponds to the standard deviation within $\pm 0.1\, \rm{\micron}$ from $\lambda$). Note that uncertainties in the model fluxes do not include the calibration errors, which are expected to be of about $20$--$30\%$.}\label{tab_midi}
   \centering
   \begin{tabular}{c | c c c c | c}
      \hline
      $\lambda$   & FWHM  & FWHM & Model flux & Red. $\chi^2$  & VISIR spec. \\
      \rm{\micron} & mas   & pc & mJy         &                   & mJy    \\
      \hline
       $8.5$       & $< 3.9$ & $< 0.3$ & $42.2^{+3.3}_{4.0}$    & $0.51$   & $53  \pm 3$  \\[0.1cm]
      $10.5$       & $< 5.7$ & $< 0.5$ & $88.7^{+8.8}_{13.2}$   & $0.12$   & $100 \pm 7$  \\[0.1cm]
      $12.0$       & $< 6.8$ & $< 0.6$ & $117.2^{+12.8}_{12.2}$ & $0.93$   & $123 \pm 7$  \\[0.1cm]
      \hline
   \end{tabular}
\end{table}

The best-fit parameters obtained for the Gaussian model fit are listed in Table\,\ref{tab_midi} along with their corresponding $1 \sigma$ uncertainties. The radial plots of the best-fit model are shown in Fig.\,\ref{fig_radial}, together with the associated $1 \sigma$ uncertainty region for each wavelength band. Overall, the fit suggest that the nucleus of NGC\,1052 is an unresolved source even at the longest baselines. For the three bands, the total flux of the unresolved emission is well constrained, the intersection at zero-metre baseline length is found inside the $1 \sigma$ error flux of the single-dish measurements at $10.5\, \rm{\micron}$ and $12\, \rm{\micron}$, while at $8.5\, \rm{\micron}$ the single-dish flux is slightly above the uncertainty but still compatible with the correlated flux at 130\,m. On the other hand, the Gaussian FWHM is unconstrained and thus the values shown in Table\,\ref{tab_midi} shall be considered as upper limits to the size of the brightness distribution. At $10.5\, \rm{\micron}$ and $12.0\, \rm{\micron}$ the best-fit model is somewhat compatible with a marginally resolved source. However, most of the mid-IR nuclear emission in NGC\,1052 is confined within a very compact region of $< 5\, \rm{mas}$ in size. At a distance of $18\, \rm{Mpc}$, this corresponds to a size smaller than $< 0.5\, \rm{pc}$.

\subsection{Power-law dust distribution}\label{plaw}
A more physically motivated model was also applied to infer the dust size distribution, following the approach described in \citet{2007A&A...476..713K} and \citet{2011A&A...536A..78K}. The model assumes a power-law dust density distribution ($\propto r^\gamma$) down to a certain inner radius ($R_{\rm in}$) where the profile is truncated due to the sublimation of the dust grains by the AGN radiation. We fixed the inner radius at $R_{\rm in} = 9.4\, \rm{light\,days}$, which corresponds to the expected value derived from the correlation between the accretion disc luminosity ($L_{\rm UV}$) and the \textit{K}~band reverberation measurements obtained for nearby Seyfert 1 galaxies \mbox{\citep{2006ApJ...639...46S,2007A&A...476..713K}}. Note that this radius is, at the same AGN luminosity, about 3 times lower when compared to the \citet{1987ApJ...320..537B} sublimation radius \citep{2009A&A...493L..57K}. The $L_{\rm UV}$ estimate is based on the $L_{\rm UV}$--$L_{\rm X}$ relation from \citet{2012A&A...539A..48M}, using a $2$--$10\, \rm{keV}$ absorption-corrected luminosity\footnote{Adapted to the distance used in this work.} of $L_{\rm X} \sim 4.4 \times 10^{41}\, \rm{erg\,s^{-1}}$ \citep{2009ApJ...698..528B}, which translates into $L_{\rm UV} \sim 2.8 \times 10^{42}\, \rm{erg\,s^{-1}}$. Additionally, we assumed a dust sublimation temperature of $1500\, \rm{K}$, a grain size of $0.05\, \rm{\micron}$, and a temperature gradient index of $\beta \sim -0.36$ for the dust in the torus, assuming standard interstellar-medium dust grains ($T \propto r^\beta$; \citealt{1987ApJ...320..537B}; \citealt{2009A&A...493L..57K}).

\begin{figure}
   \centering
   \includegraphics[width=1.05\linewidth]{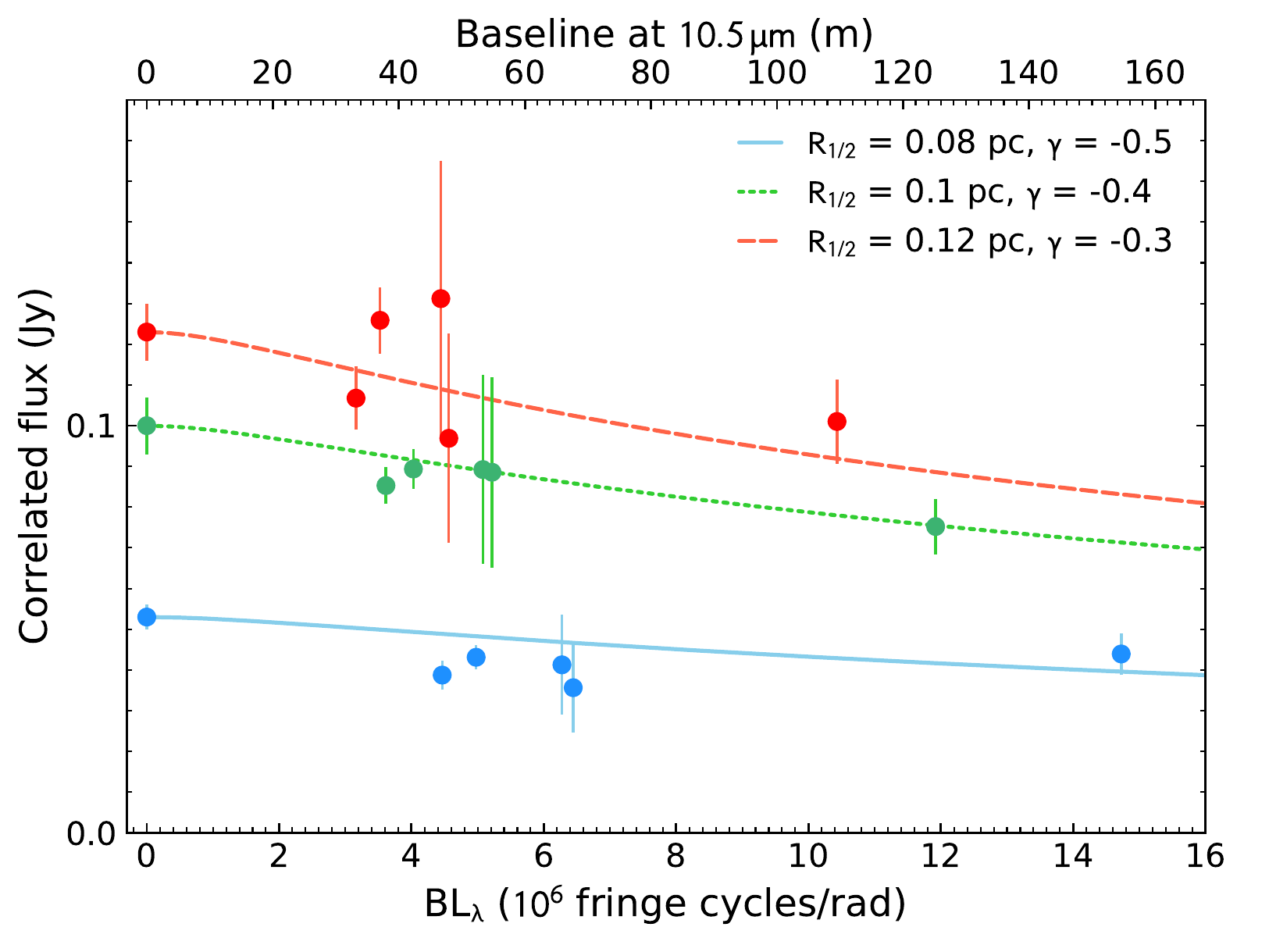}
   \caption{Best-fit power-law models to the MIDI correlated fluxes vs. baseline projected length for the nucleus of NGC\,1052. The MIDI measurements have been grouped in three wavelength bands centred at $8.5\, \rm{\micron}$ (blue-solid line), $10.5\, \rm{\micron}$ (green-dotted line), and $12.0\, \rm{\micron}$ (red-dashed line). Values obtained for the half-light radius ($R_{1/2}$) and the dust density profile index ($\gamma$) are indicated in the legend.}\label{fig_plaw}
\end{figure}

For the three wavelength bands extracted from MIDI, a steep dust density profile ($\gamma \sim -0.3$ to $-0.5$), typical of bright AGN \citep{2011A&A...536A..78K}, has to be assumed for the three bands in order to reproduce the correlated fluxes (Fig.\,\ref{fig_plaw}). A more extended brightness distribution would have been clearly resolved with the widest baselines. Even for the longest wavelengths, the half-light radius is $R_{1/2} = 0.12\, \rm{pc}$, i.e. only about 15 times the size of the inner radius. If a lower X-ray luminosity is adopted, e.g. $L_{\rm X} \sim 1.4 \times 10^{41}\, \rm{erg\,s^{-1}}$ \citep{2009A&A...506.1107G}, a smaller inner radius of $R_{\rm in} = 4.8\, \rm{light\,days}$ and a flatter dust density distribution with $\gamma \sim -0.1$ would result in a similar $R_{1/2}$ value, since the radial brightness distribution would be then dominated by the temperature profile in this case. If the limit for large grains is considered ($\beta = -0.5$), the brightness distribution at $12\, \rm{\micron}$ would not be marginally resolved unless an increasing dust density profile with increasing radius ($\gamma \gtrsim 1$) is adopted. Overall, the results obtained with the power-law model are in agreement with the Gaussian analysis, i.e. \textit{if dust emission represents a significant contribution to the mid-IR nuclear flux in NGC\,1052, this must be confined within the innermost few tenths of parsec (less than 6 sublimation radii)}.

\subsection{IR polarisation}\label{pol_results}
In Table\,\ref{tab_pol} we compile the IR to UV polarisation measurements for the nucleus of NGC\,1052, including our measurements and values published in the literature. The first optical polarimetric observations of NGC\,1052 were presented by \citet{1973ApJ...179L..93H} in the UV range, reporting a linear polarisation degree of $3.28 \pm 0.45\%$ at PA\,$= 42\degr$ within a $10\farcs2$ aperture diameter in the \textit{U}~band, and \citet{1982ApJ...252L..53R} in the \textit{K}~band, who measured a $0.81 \pm 0.16\%$ degree at PA\,$= 140\degr$ within a $7\farcs8$ beam (i.e. an intrinsic degree of $4.5\%$, after the subtraction of the unpolarised stellar light contribution). Note that the angle of the linear polarisation changes about $60\degr$--$80\degr$ from optical to IR wavelengths.

The polarisation degree measured in our two \textit{N}~band filters are in agreement within the uncertainties. In contrast, the polarisation degree measured in the \textit{K}~band is a factor of $\sim 8$--$10$ lower, while its PA differs by $\sim 45\degr$--$50\degr$ when compared to that in the \text{N}~band (Table\,\ref{tab_pol}). Since the \textit{K}~band polarisation degree may be seriously affected by the contribution of unpolarised stellar light, one can estimate the intrinsic degree of polarisation by comparing the nuclear fluxes using a $0\farcs75$ aperture radius and those obtained at the highest angular resolution ($0\farcs15$), both measured in the NaCo image (see Tables\,\ref{sed_low} and \ref{sed_high}, respectively). The former includes a factor of $\sim 10.6$ higher flux, thus considering this dilution factor and assuming that all the polarised radiation is originated in the unresolved core, an intrinsic linear polarisation degree of $4.8 \pm 0.7 \%$ in the \textit{K}~band is estimated. This would be in agreement with the intrinsic $4.5\%$ derived by \citet{1982ApJ...252L..53R}, albeit the difference in the PA is still large ($29\degr$). These authors obtain the intrinsic value from the observed degree of $0.81\%$ within a $7\farcs8$ beam, after subtracting the underlying galaxy. Note that our LIRIS measurement of $0.45\%$ within a diameter of $1\farcs5$ has been corrected for the polarisation bias, while \citet{1982ApJ...252L..53R} do not mention such correction in their reported value. This could explain the difference in the observed polarisation degree and PA, and the agreement in the intrinsic degree derived in both cases.
\begin{table}
  \caption{Linear polarisation measurements for the nucleus of NGC\,1052 in the IR to UV range. Note that \citet{1982ApJ...252L..53R} derive an intrinsic polarisation degree of $4.5\%$ in the \textit{K}~band from the measured $0.81\%$ after the subtraction of the unpolarised stellar light.}\label{tab_pol}
  \centering
  \begin{tabular}{c c c c c c}
      \hline
      Band & $\lambda_c$ &    Aper. & $P_{\rm lin}$ &  PA & Ref.\\
           &     \micron &   radius &          \%  & deg & \\
      \hline
      \textit{U}  & $0.38$ & $5\farcs1$  & $3.28 \pm 0.45$ & $42$ & \citealt{1973ApJ...179L..93H} \\[0.1cm]
      \textit{Ks} & $2.15$ & $0\farcs75$ & $0.45 \pm 0.07$ & $169 \pm 9$ & This work \\[0.1cm]
      \textit{K}  & $2.2$  & $3\farcs9$  & $0.81 \pm 0.16$ & $140$ & \citealt{1982ApJ...252L..53R} \\[0.1cm]
      Si2         & $8.7$  & $0\farcs5$  &   $4.4 \pm 0.5$ & $125 \pm 4$ & This work \\[0.1cm]
      Si5         & $11.6$ & $0\farcs5$  &   $3.5 \pm 0.6$ & $118 \pm 5$ & This work \\[0.1cm]
      \hline
  \end{tabular}
\end{table}

\begin{figure}
   \centering
   \includegraphics[width=0.9\columnwidth]{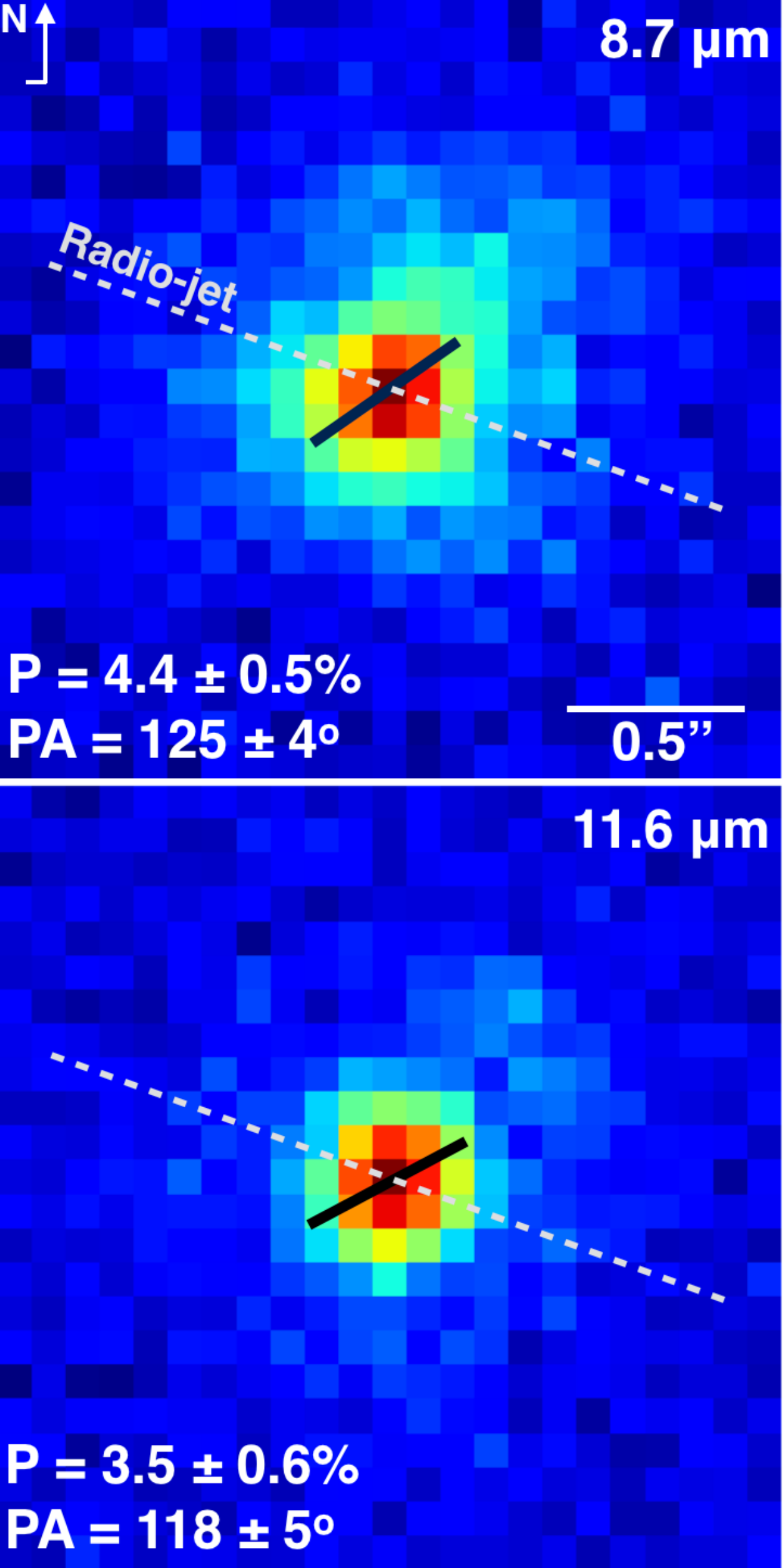}
   \caption{$8.7\, \rm{\micron}$ (top) and $11.6\, \rm{\micron}$ (bottom) continuum images of NGC\,1052 taken with CanariCam on the 10.4-m GTC. The nuclear polarisation vector (black line) with their values (bottom left of each plot) in a $0\farcs5$ ($42.5\, \rm{pc}$) aperture are shown. The radio-jet (grey dashed line) with a PA\,$ = 70\degr$ \citep{2004A&A...426..481K} is shown. North is up and East is to the left.}
   \label{fig_ngc1052_CCpol}
\end{figure}

Figure \ref{fig_ngc1052_CCpol} shows the total intensity maps in the $8.7$ and $11.6\, \rm{\micron}$ filters with the corresponding nuclear polarisation vector overplotted (black line). In both filters NGC\,1052 appears as a point-like source with a very faint extended tail to the NW. The latter however is not detected in the images by VISIR or T-ReCS. The angular separation between the radio jet axis at PA\,$ = 70 \degr$ (white-dashed line; \citealt{2004A&A...426..481K}) and the IR polarisation angle is of about $35\degr$ in the \textit{K}~band and $\sim 50\degr$ in the \textit{N}~band, and could be caused by Faraday rotation of the polarised radiation at radio wavelengths \citep{2004A&A...426..481K}, similar to the case of M87 \citep{2016ApJ...832....3A}. A polarisation angle perpendicular to the direction of the magnetic field is expected for optically-thin synchrotron radiation, while optically-thick emission shows instead a parallel alignment \citep{1970ranp.book.....P,2014ApJ...780...87M}. When compared with previous mid-IR polarimetric measurements of AGN, NGC\,1052 shows a relatively high polarisation degree \citep[see review by][]{2018MNRAS.478.2350L}, only exceeded by Cygnus\,A \citep{2018ApJ...861L..23L}, where the polarisation is ascribed to dichroic absorption and emission by aligned dust grains.

\section{Discussion}\label{discuss}

In this Section we will investigate the two main scenarios introduced in Section\,\ref{intro} to interpret the observations of NGC\,1052. For each scenario we will address three essential aspects to understand the properties of this nucleus: the size of the mid-IR brightness distribution in the context of previous results for Seyfert nuclei, the origin of the polarised radiation, and the evidences in favour and against the presence of a nuclear absorber. Finally, in Section\,\ref{compactjet} we will analyse the shape of the nuclear flux distribution, the measured optical\,--\,soft X-ray index, and the location of NGC\,1052 in the mid-IR\,--\,X-ray correlation.

\subsection{A very compact dust distribution}\label{compactdust}
Since the mid-IR brightness distribution in NGC\,1052 is confined within the innermost $\lesssim 0.5\, \rm{pc}$, the nuclear dust distribution cannot be resolved using the longest baselines available for the UTs ($\sim 130\, \rm{m}$). This is in contrast with other nearby AGN studied by MIDI, which show extended dust distributions (e.g. \citealt{2014A&A...565A..71L}). For a proper comparison, we estimated the sizes predicted by different calibrations based on the $12\, \rm{\micron}$ nuclear luminosity. Assuming the calibration provided by \citet{2011A&A...531A..99T}, for a $L_{\rm 12 \mu m} = 120\, \rm{mJy}$ (Table\,\ref{tab_midi}) a FWHM size of $0.5 \pm 0.1\, \rm{pc}$ is expected, in marginal agreement with our upper limit of $< 0.5\, \rm{pc}$. The differences are however larger when NGC\,1052 is compared with the results obtained by \citet{2011A&A...536A..78K} in their AGN sample, where lower luminosity nuclei tend to show comparatively more extended dust distributions. Specifically, they find $12\, \rm{\micron}$ half-light radii to be nearly constant around $R_{1/2} \sim 1\, \rm{pc}$, almost independently of the nuclear luminosity ($\propto L^{0.01}$). Following the approach in \mbox{\citet{2011A&A...536A..78K}} we derive $R_{1/2} \sim 0.12\, \rm{pc}$ for NGC\,1052, i.e. nearly one order of magnitude smaller than expected.

\begin{figure*}
   \centering
   \includegraphics[width=0.9\hsize]{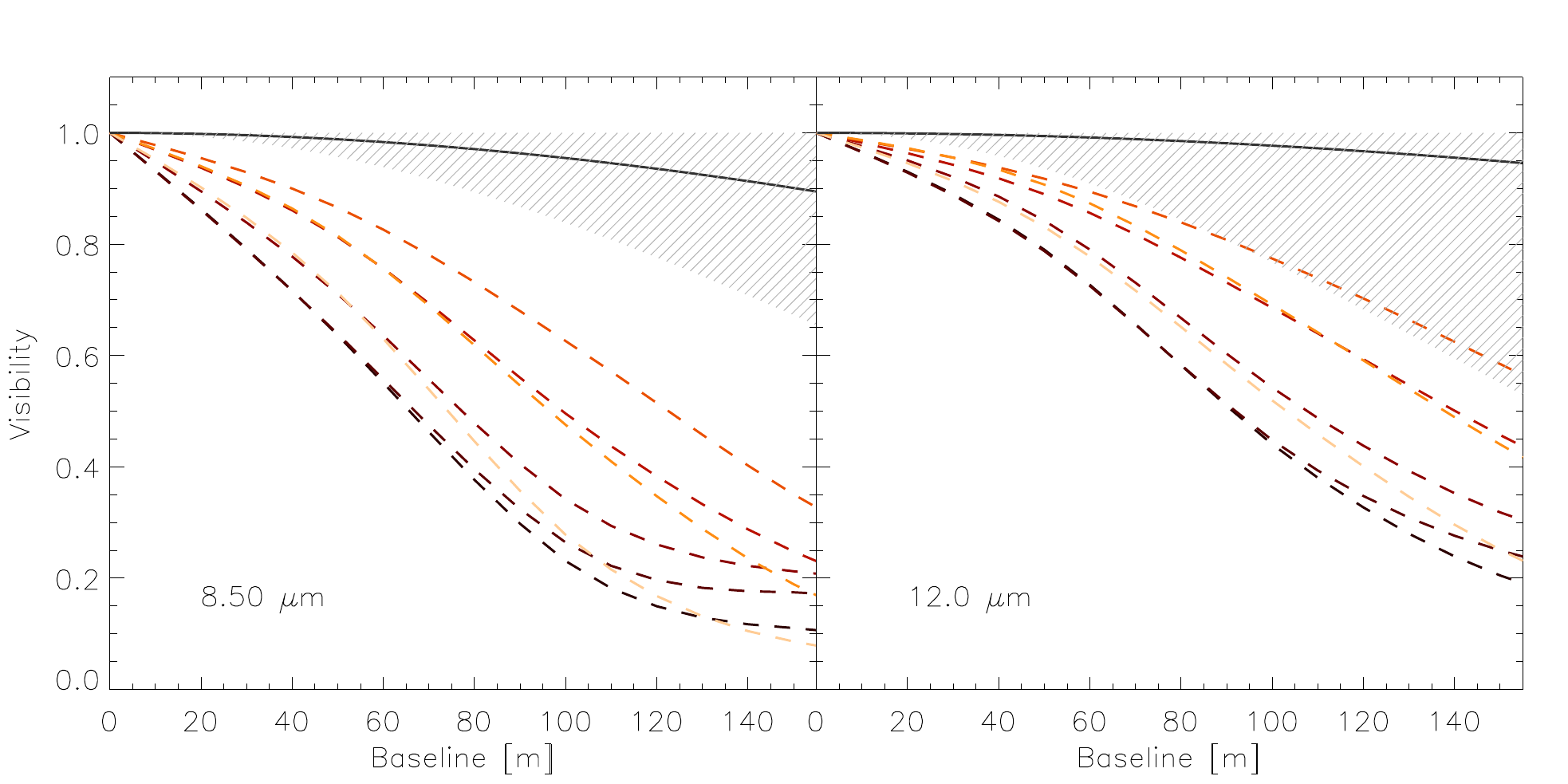}
   \caption{Visibility distribution of the Gaussian model used to fit the mid-IR correlated fluxes in NGC\,1052 (black line) and the Circinus Galaxy (dashed lines; \citealt{2014A&A...563A..82T}). The latter has been re-scaled to the distance and $12\, \rm{\micron}$ luminosity of NGC\,1052 following the $r \propto \sqrt{L}$ trend derived by \citet{2011A&A...531A..99T} for a sample of Seyfert nuclei. The different colours of the dashed lines represent visibilities measured along different position angles, from the most extended axis along PA\,$ = 137\degr$ (black-dashed line) to the most compact one along $17\degr$ (orange-dashed line). The scaled brightness distribution in Circinus is only marginally compatible with NGC\,1052 at $12\, \rm{\micron}$ along the most compact direction.}
   \label{fig_cir-n1052}
\end{figure*}

These significant differences among the existing calibrations might be indicative of the lack of a common size--luminosity relation, as suggested by \citet{2013A&A...558A.149B}. Nevertheless, a short experiment can still be performed to understand how NGC\,1052 compares to the Circinus galaxy, which has been extensively studied using MIDI and is currently the object with the best $(u,v)$ plane coverage in the mid-IR \citep{2014A&A...563A..82T}. NGC\,1068 has also a well sampled $(u,v)$ plane, however the contribution of multiple shock regions located along the jet dominates the nuclear morphology in the IR, which is not shaped only by the torus emission (\mbox{\citealt{2014A&A...565A..71L}}). Moreover, the system axis of the central AGN has a high inclination angle in both NGC\,1052 ($\sim 72\degr$; \citealt{2004A&A...426..481K,2016A&A...593A..47B}) and Circinus ($\gtrsim 65\degr$; \mbox{\citealt{2003ApJ...590..162G}}, \citealt{2016ApJ...832..142Z}), i.e. the tori in these galaxies are supposed to show a nearly edge-on orientation with respect to our line of sight. Furthermore, this experiment allows us to compare the brightness distributions at shorter wavelengths, since the estimates mentioned above refer to the expected size at $12\, \rm{\micron}$.

Thus, the mid-IR brightness distribution of the Circinus Galaxy has been re-scaled to match the distance and luminosity. Fig.\,\ref{fig_cir-n1052} shows the radial visibility distribution at $8.5\, \rm{\micron}$ and $12\, \rm{\micron}$ of the best-fit Gaussian models obtained for NGC\,1052 (solid line) and Circinus at different baseline orientations (dashed lines; see fig.\,5 in \mbox{\citealt{2014A&A...563A..82T}}), where $17\degr$ (black-dashed line) and $137\degr$ (orange-dashed line) correspond to the most compact and the most extended axes in this nucleus, respectively. The size values for the Circinus models at different orientations have all been re-scaled by a factor of $0.5$, i.e. following the size--luminosity trend at $12\, \rm{\micron}$ by \citet{2011A&A...531A..99T} ($r \propto \sqrt{L}$; assuming a $12\, \rm{\micron}$ flux of $\sim 15$--$20\, \rm{Jy}$ at $4\, \rm{Mpc}$ and $0.18\, \rm{Jy}$ at $18\, \rm{Mpc}$ for Circinus and NGC\,1052, respectively). If instead the wavelength-dependent scaling relation found by \citet{2011A&A...536A..78K} is adopted, the size correction factors for the case of Circinus would be of about $0.7$ at $8.5\, \rm{\micron}$ and $\sim 1$ at $12\, \rm{\micron}$. In this work we adopted the trend obtained by \citet{2011A&A...531A..99T} since this scenario predicts a more compact dust distribution, and thus closer to the case of NGC\,1052. The left panel in Fig.\,\ref{fig_cir-n1052} shows that NGC\,1052 is significantly smaller than the scaled brightness distribution of Circinus at $8.5\, \rm{\micron}$, even along its more compact dimension. At $12\, \rm{\micron}$ the two distributions would be marginally in agreement only along the most compact axis in Circinus (right panel), in line with the size estimated above using the correlation in \citet{2011A&A...531A..99T}. If the two distributions were comparable, a large drop in the correlated fluxes would have been expected for any baseline orientation at $8.5\, \rm{\micron}$ and for most of them at $12\, \rm{\micron}$. That is, a scaled version of the dust distribution in Circinus would have been clearly resolved by MIDI in the nucleus of NGC\,1052. Therefore, from the comparison with the different torus radial structures described in \citet{2011A&A...531A..99T} and \citet{2011A&A...536A..78K}, and the scaling of the dust distribution in Circinus, \textit{we conclude that assuming a nucleus dominated by thermal emission in the mid-IR for NGC\,1052 would imply an extremely compact radial dust distribution not seen in other AGN}.

\begin{figure*}
   \centering
   \includegraphics[width = 0.5\textwidth]{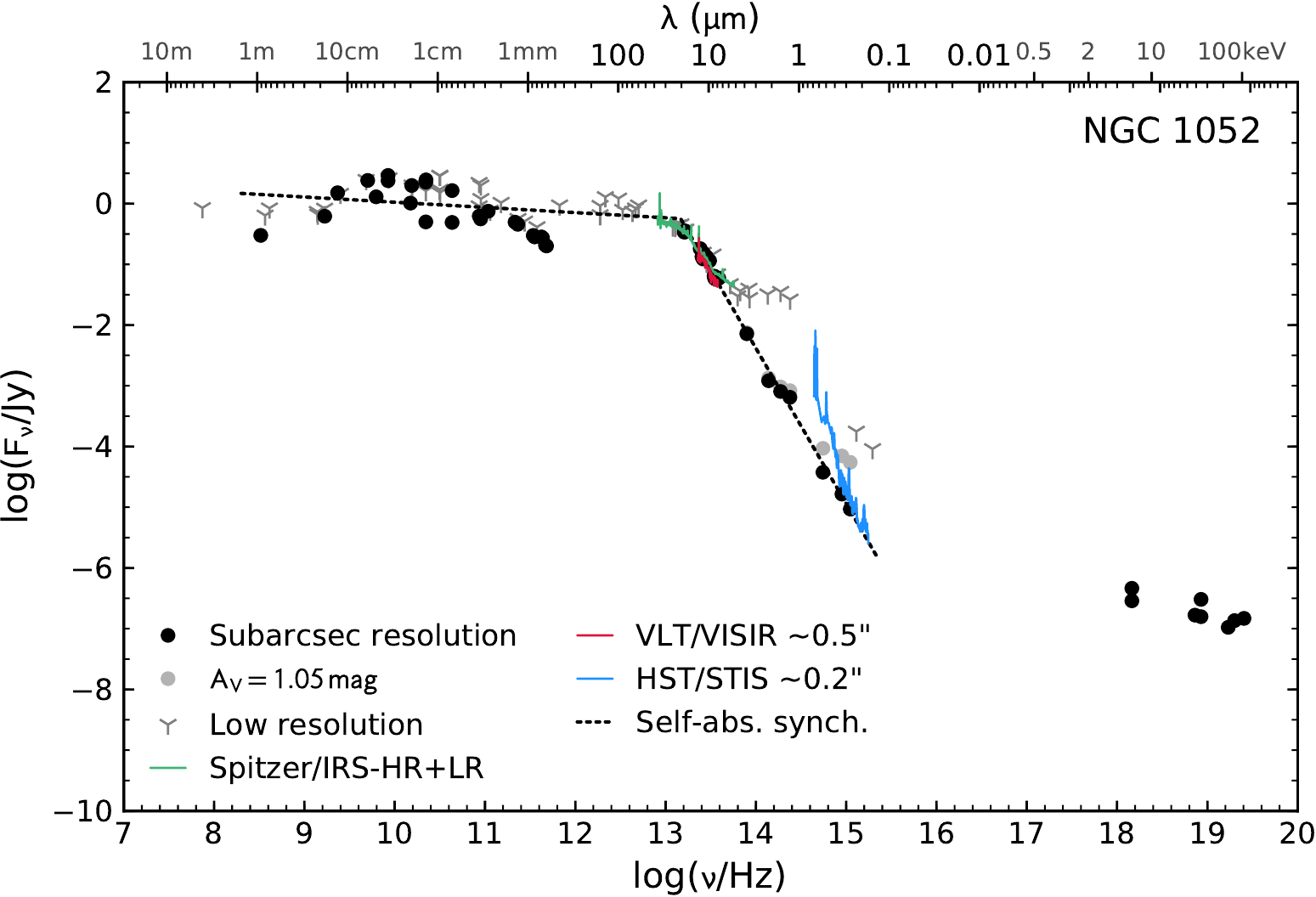}~
   \includegraphics[width = 0.5\textwidth]{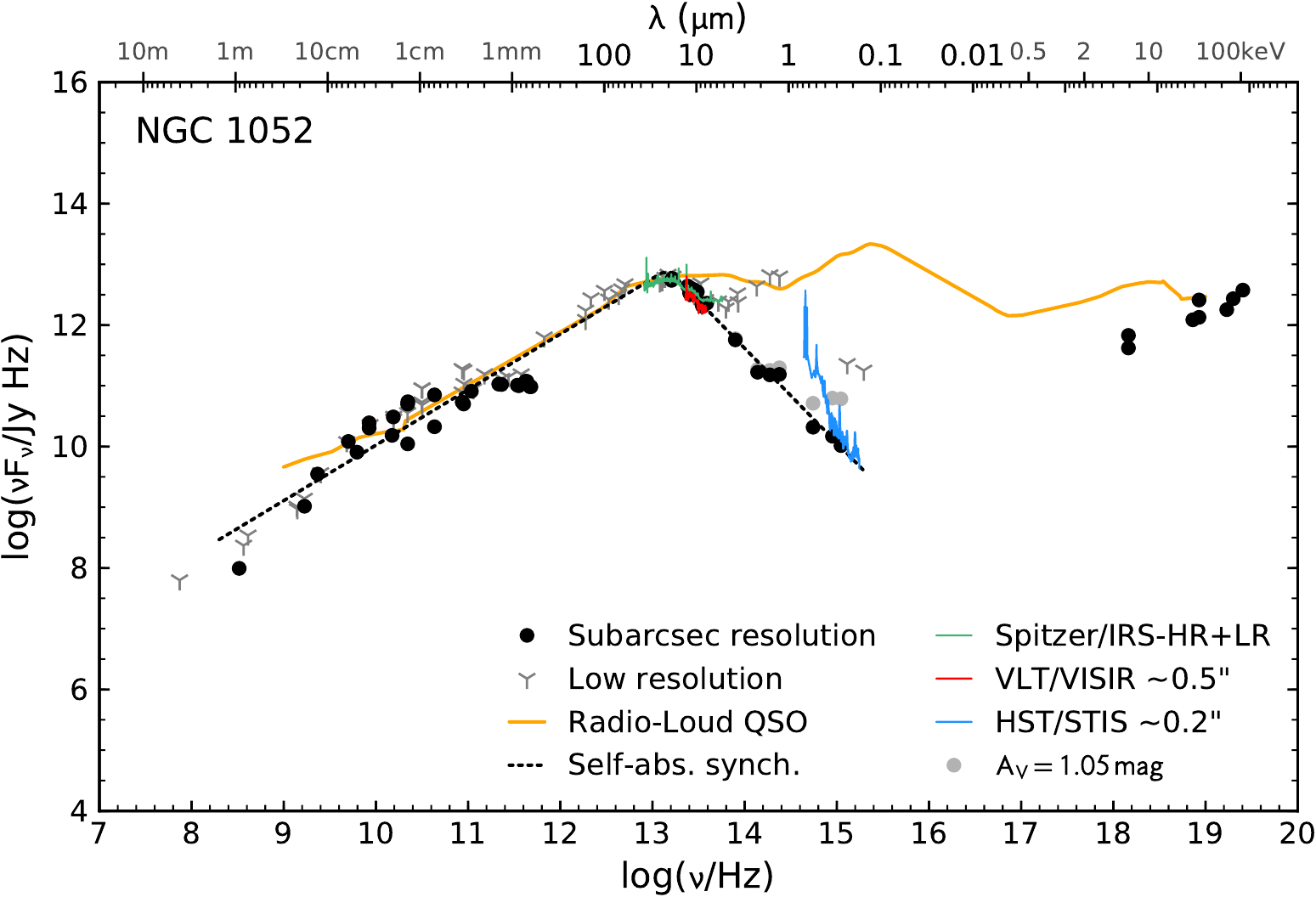}
   \caption{Rest-frame spectral flux (left panel) and spectral energy (right) distributions for the nucleus of NGC\,1052 compiled for this work from high-angular resolution data ($\lesssim 0\farcs4$; black dots; see Table\,\ref{sed_high}), and low-angular resolution measurements ($> 1''$; grey spikes; see Table\,\ref{sed_low}). The $0\farcs2$ slit width spectrum from \textit{HST}/STIS indicates possible variability in the optical range or galaxy starlight contamination (in blue). Nuclear fluxes corrected by an extinction of $A_V = 1.05\, \rm{mag}$ \citep{2015ApJ...801...42D} are shown as grey circles. Flux errors are smaller than the plotted symbols and thus are not shown. The agreement among the mid-IR spectra from \textit{Spitzer}/IRS-HR+LR ($3\farcs7 \times 57''$ aperture in the $5.2$--$14.5\, \rm{\micron}$ range, $4\farcs7 \times 11\farcs3$ aperture in the $10$--$19\, \rm{\micron}$ range, $11\farcs1 \times 22\farcs3$ aperture in the $19$--$37\, \rm{\micron}$ range; in green), VLT/VISIR ($0\farcs52$ slit width extraction; in red), and the subarcsec resolution photometry suggests that the nuclear emission dominates the energy output of the galaxy longwards of $\gtrsim 8\, \rm{\micron}$. The dotted line shows a broken power-law fit to the high-angular resolution data. This model should be considered only as a first order approximation of the flux distribution, since the radio-to-millimetre continuum is associated with optically-thick self-absorption synchrotron emission. In this scenario, the ripples observed below $10^{12}\, \rm{Hz}$ are ascribed to optical depth effects and variability (see \citealt{2019arXiv190102639B}). For comparison, the average radio-loud quasar template from \citet{1994ApJS...95....1E} is also shown on the right panel (yellow-solid line).}\label{fig_sed}
\end{figure*}

An important aspect that a compact dust distribution should explain is the degree of polarisation measured in the IR, which would be associated with scattering and/or dichroic absorption and emission by dust. Following the Serkowski's law, a $4.8\%$ degree of linear polarisation in the \textit{K}~band would imply an intrinsic degree of $\sim 24\%$ in the \textit{U}~band \citep{1975ApJ...196..261S,1996ASPC...97..125W}. However, the nuclear flux is about $\gtrsim 1.5$ orders of magnitude fainter when compared to the integrated continuum in the \textit{U}~band within the inner $\sim 10''$ (see Fig.\,\ref{fig_sed}), suggesting that the $3.3\%$ measured by \citep{1973ApJ...179L..93H} are not caused by the nuclear polarisation. On the other hand, dust scattering would have a minor contribution in the mid-IR, where the dominant polarisation mechanism are dichroic absorption and emission \citep[e.g.][]{2018ApJ...861L..23L}. Following the relation between the degree of linear polarisation produced by the dichroic mechanism and the optical depth given by \citet{1989ApJ...346..728J}, a $4.8\%$ degree in the \textit{K}~band would correspond to an opacity of $\tau_K \sim 2.8$, that is $A_V \sim 26\, \rm{mag}$ assuming the extinction curve from \citet{1989ApJ...345..245C}. Even under the maximum dichroic efficiency assumption (e.g. \citealt{2014MNRAS.444..466R}), at least $A_V \gtrsim 2.8\, \rm{mag}$ would be required to reproduce the \emph{K}~band polarisation.

An extinction of $A_V \sim 26\, \rm{mag}$ would be in agreement with the presence of a dusty torus. This is also in line with the high column density of $N_{\rm H} \sim 10^{23}\, \rm{cm^{-2}}$, associated with an X-ray absorbent medium, i.e. ionised and/or neutral gas. However, three important facts point against serious absorption affecting the optical/UV range in NGC\,1052: \textit{i)} the nucleus is detected up to $2000$\,\AA \ tightly following a power-law continuum with no trace of heavy absorption (Fig.\,\ref{fig_sed}), which would determine the shape of the spectrum in this range; \textit{ii)} the $A_V = 1.05\, \rm{mag}$ measured in the \textit{HST}/FOS spectrum by \citet{2015ApJ...801...42D}; and \textit{iii)} the detection of the silicate band in emission (Fig.\,\ref{fig_irspec}), which is hard to accommodate for a nearly edge-on system. Additionally, although the presence of a broad H$\alpha$ lines in polarised light \citep{1999ApJ...525..673B} proves the existence of a scattering medium, the detection of broad H$\alpha$ and H$\beta$ in the unpolarised spectrum by \citet{2005ApJ...629..131S} also suggest a direct line of sight to the innermost nuclear region. On the other hand, NGC\,1052 shows a variable N$_{\rm H}$ \citep{2014A&A...569A..26H} together with a ``harder when brighter'' behaviour in X-rays \citep{2016MNRAS.459.3963C}. This means that the X-ray spectrum becomes steeper with increasing luminosity, while the opposite is expected, e.g. if one associates a larger N$_{\rm H}$ with an increase in the optical depth of the absorber along the line of sight.

In this context, the silicate emission feature could originate close to the active nucleus in an optically-thin dusty wind (e.g. \citealt{2012ApJ...749...32K}), or be associated with the compression of the ISM by jet-driven shocks as revealed by the detection of water masers along the radio-jet axis \citep{1998ApJ...500L.129C}. Alternatively, the tentative decrease in the correlated fluxes at $10.5$ and $12\, \rm{\micron}$ could be indicative of extended emission on scales larger than the mid-IR core ($\gtrsim 0.5\, \rm{pc}$), while the nucleus remains compact at $8.5\, \rm{\micron}$. For instance, extended $9.7\, \rm{\micron}$ silicate emission in elliptical galaxies with similar strengths as in NGC\,1052 has been previously reported by \citet{2006ApJ...639L..55B}, and is associated with the warm dust produced by mass-losing evolved stars.

\subsection{A compact jet}\label{compactjet}
An alternative explanation to the compactness of the mid-IR emission is the presence of a strong non-thermal component, possibly associated with an unresolved jet. To investigate this scenario we consider the high-angular resolution spectral energy distribution (black dots in Fig.\,\ref{fig_sed}), along with the low-angular resolution fluxes that include the host galaxy contribution (grey spikes in Fig.\,\ref{fig_sed}). Nuclear fluxes swing up and down in the $0.3$--$3\, \rm{Jy}$ range from $\sim 3 \times 10^8\, \rm{Hz}$ to $100\, \rm{GHz}$. Then they reach a turnover at $\sim 20\, \rm{\micron}$ ($1.5 \times 10^{13}\, \rm{Hz}$), followed by a well defined power-law continuum with an index of $\alpha = 2.58 \pm 0.02$ ($S_\nu \propto \nu^{-\alpha}$; \citealt{2015ApJ...814..139K}; dotted line in Fig.\,\ref{fig_sed}). The presence of such a steep continuum was predicted in the pioneering work of \citet{1982ApJ...252L..53R}. This shape departs largely from the big blue bump that dominates the optical/UV range in quasars (yellow-solid line in Fig.\,\ref{fig_sed}, right panel; \citealt{1994ApJS...95....1E}), associated with the accretion disc emission. X-ray fluxes depart clearly for the extrapolation of the IR-to-UV power law, and are likely associated with inverse Compton radiation \citep{2018MNRAS.478L.122R}.

Longwards of $\sim 8\, \rm{\micron}$, the total flux is completely dominated by the nuclear emission, as suggested by the agreement between low- and high-angular resolution measurements. In the optical and near-IR ranges the galaxy starlight has a major contribution and only the subarcsec resolution data allow us to recover the nuclear fluxes. Note that the datasets used were acquired over several decades, in particular subarcsec IR-to-UV fluxes span a period of $18\, \rm{yr}$. However they show a remarkably consistent shape over two orders of magnitude in frequency. Only the \textit{HST}/STIS spectrum, extracted from a $0\farcs2$ slit width, shows a continuum level about $8$ times higher than the flux measured in the F555W band. This could be ascribed to possible contamination by the underlying starlight, which has not been subtracted from the nuclear STIS spectrum. We also note that F555W was acquired with the Wide-Field Planetary Camera (WFPC1), thus it is affected by the spherical aberration. If we correct the F555W flux to a similar encircled energy fraction as that expected for the other \textit{HST} bands ($60$--$80\,\%$), the flux would increase by a factor of 3-4. Thus, the host galaxy contribution in the STIS spectrum and the light spread due to the aberration in the F555W image might explain the difference in the continuum flux between both observations.

The nuclear continuum distribution in Fig.\,\ref{fig_sed} reminds closely to the characteristic pattern of self-absorbed synchrotron emission, i.e. the signature of non-thermal radiation produced by a compact jet \citep{1979ApJ...232...34B}. The radio emission shows a somewhat flat distribution associated with optically-thick radiation, followed by a turnover and a decreasing power-law continuum with increasing frequency in the optically-thin range. Note that the power-law fit in the optically-thick domain shown in Fig.\,\ref{fig_sed} should be considered only as a first order approximation, since the ripples observed in the flux distribution at frequencies $< 100\, \rm{GHz}$ are likely caused by the superposition of different jet components and also by flux variability (\citealt{1983ApJ...267L..11H}, \citealt{2007A&A...469..451T}). A four year VLBI monitoring program ($2005$--$2009$) at $22$ and $43\, \rm{GHz}$ recently published by \citet{2019arXiv190102639B} reveals large amplitude variations of about $1.8\, \rm{Jy}$ ($78 \%$) and $1.2\, \rm{Jy}$ ($70 \%$), respectively, for the total nuclear emission (see Table\,\ref{sed_high}), which can account for the apparent bump seen in the high-angular resolution data at radio wavelengths (Fig.\,\ref{fig_sed}). Optical depth effects can seriously affect the continuum shape in this range, thus we keep a simple power law in our fit since capturing the complex distribution of the optically-thick components in the jet is beyond the scope of this paper. A combination of IR dust and radio synchrotron emission can also explain the overall flux distribution of radio galaxies \citep[e.g.][]{2004A&A...426L..29M,2018ApJ...861L..23L}. However in the case of NGC\,1052 the VLBI radio core, which suffers from heavy free-free absorption below $\lesssim 43\, \rm{GHz}$ \citep{2004A&A...426..481K}, becomes the dominant nuclear contribution ($\sim 70\%$ of total flux) at $86\, \rm{GHz}$ (Tables\,\ref{sed_high} and \ref{sed_low}; \citealt{2016A&A...593A..47B}). This flux is confined within the innermost $0.407 \times 0.075\, \rm{mas^2}$ region ($42.3 \times 7.8\, \rm{light\,days^2}$), that is of the order of the $R_{\rm in}$ estimated in Section\,\ref{plaw}, suggesting that most of millimetre/submillimetre emission comes from the vicinity of the black hole\footnote{$\sim 450$--$2400\, \rm{R_S}$ in Schwarzschild radii for $M_{\rm BH} = 1.55 \times 10^8\, \rm{M_\odot}$.}. The ALMA archival observations at $478\, \rm{GHz}$ presented in this work, with an angular resolution of $0\farcs12 \times 0\farcs10$ ($10 \times 9\, \rm{pc^2}$), do not show any extended structure in the millimetre/submillimetre range that could be indicative of extended cold dust emission from the torus. This is further confirmed by recent band 7 observations at $65 \times 53\, \rm{mas}$ resolution ($5.7 \times 4.6\, \rm{pc}$; Kameno et al. priv. comm.). On the contrary, the ALMA fluxes included in Table\,\ref{sed_high} are in good agreement with large aperture measurements in Table\,\ref{sed_low}. Future continuum ALMA observations at sub-parsec resolution ($\lesssim 0\farcs01$) at higher frequencies ($\gtrsim 600\, \rm{GHz}$) would be useful to probe the possible contribution of thermal cold dust emission. Besides the compact emission, the existence of an extended jet at radio wavelengths is well know on both arcsecond \citep{1984ApJ...284..531W, 2007ApJS..171..376C} and milliarcsecond scales \citep{1998ApJ...500L.129C,2019arXiv190102639B}.

Regarding the polarisation, a compact jet does not require any extinction to explain the degree of linear polarisation measured in this work. The difference in the polarisation angle between the near- and the mid-IR measurements ($\sim 50\degr$; Table\,\ref{tab_pol}) suggests that multiple jet components might be included within the inner few tens of parsecs. This is the case of M87, where abrupt changes of $\sim 90\degr$ have been found between contiguous regions resolved along the jet, likely driven by the helical structure of the magnetic field \citep{2016ApJ...832....3A}. Since the relative contribution of these components to the total flux varies with wavelength --\,for instance HST-1 outshines the core in the UV at certain epochs\,-- changes in the PA of the integrated polarised flux between different filters are not surprising.

As mentioned in Section\,\ref{compactdust}, the analysis of \textit{Swift} observations revealed a ``harder when brighter'' variability behaviour in the X-rays \citep{2016MNRAS.459.3963C}, as opposed to the ``softer when brighter'' behaviour shown by Seyfert galaxies (e.g. \citealt{2012A&A...537A..87C}, \citealt{2018MNRAS.481.3563P}). This is hard to explain in terms of a variable column gas density, since a harder X-ray slope produced by an increasing N$_{\rm H}$ should be followed by a drop in the observed flux. However, the opposite behaviour is a well known characteristic of X-ray binaries in the hard state. This is ascribed to the recession of the disc and the transition to an advection-dominated flow at low accretion rates \citep{1995Natur.374..623N}. At this point the self-Comptonisation of synchrotron photons radiated by the corona and/or the jet becomes the dominant contribution in the X-rays. This mechanism applies also to blazars, where the injection of highly energetic electrons in the jet, e.g. due to a shock re-acceleration or a magnetic re-connection event, increases both the overall luminosity and the peak frequency of the inverse Compton component ($10\, \rm{MeV}$--$100\, \rm{GeV}$), steepening the observed X-ray spectral slope (e.g. \mbox{\citealt{2009MNRAS.396L.105G}}, \mbox{\citealt{2018ApJ...867...68W}}). A similar trend has also been observed in the LINER nucleus of NGC\,7213 \citep{2012MNRAS.424.1327E}. In this context, we interpret the $N_{\rm H} \sim 10^{23}\, \rm{cm^{-2}}$ column density as a dense neutral and/or ionised gas absorber in the innermost nuclear region, but not associated with a dusty medium. This is in line with the detection of broad lines in both the polarised and the total intensity spectra. Additionally, NGC\,1052 fits in the correlation found by \citet{2015ApJ...814..139K} between the X-ray power-law index and the turnover frequency found at IR-to-radio wavelengths also followed by X-ray binaries, suggesting that the properties of both the plasma close to the black hole and the accelerated particles in the jet are tightly related.

A drawback of the jet scenario is that a larger variability would be expected in the IR-to-UV continuum. So far, NGC\,1052 have shown a consistent mid-IR flux over six different epochs in the 2007--2015 period, including VISIR, T-ReCS, and CanariCam, while tentative UV variability was reported by \citet{2005ApJ...625..699M}. In the millimetre range, the core displays a fairly stable flux with occasional burst (two in the 1972--1981 period; \citealt{1975ApJ...196..347H}). Thus, NGC\,1052 seems to be a relatively stable source during most of the time. A monitoring campaign in the UV range would be required in order to probe the variability associated with the most energetic particles in the jet framework.

While compact jet emission in the IR-to-optical continuum has been suggested for active nuclei in the past (e.g. M81 in \citealt{2008ApJ...681..905M}, NGC\,4151 in \citealt{2011ApJ...735..107M}), NGC\,1052 is the first LLAGN where this component has been first identified by its characteristic continuum spectral shape, and then isolated in the mid-IR range at milliarcsec resolution. The synchrotron core of Centaurus~A was studied using MIDI interferometry by \mbox{\citet{2007A&A...471..453M}} and \citet{2010PASA...27..490B}. However, in this case the shape of the continuum emission is seriously affected by the high foreground extinction ($A_V \sim 14\, \rm{mag}$), which hampers the identification of the power-law flux distribution in the optical/UV range.

Previous studies combining RIAF and jet models have been used to fit the continuum emission in LLAGN, but these efforts have not been conclusive on the nature of the UV-to-IR continuum, while the case of NGC\,1052 remains hard to reproduce by these models \citep[see][]{2009ApJ...703.1034Y,2011ApJ...726...87Y}. This is likely due to the prominent slope in the IR-to-UV range, which departs largely from the canonical optically-thin spectral index assumed in these models ($\alpha \sim 0.7$), thus RIAF plus standard jet models have serious problems to accommodate such a steep continuum. In this regard, steep optical spectra with $\alpha \sim 2$ have been found to be common among LLAGN \citep{1996ApJ...462..183H,1999ApJ...516..672H}. In the context of accelerated particles in a jet, a spectral index of $\alpha \gtrsim 2.6$ corresponds to an index of $p \gtrsim 6.2$ in the energy distribution of the accelerated particles [$N(E) dE \propto E^{-p} dE$], which would imply an extremely efficient --\,and possibly unphysical\,-- cooling process. Even if we adopt an extinction of $A_V = 1.05\, \rm{mag}$ \citep{2015ApJ...801...42D}, the corrected spectral index would be $\alpha = 2.2$ ($p = 5.4$; grey circles in Fig.\,\ref{fig_sed}). Thus, we favour a thermalised energy distribution for the radiating particles as a more likely scenario to explain the steepness of the continuum distribution. This could be associated with particles in the corona or the base of the jet that have not been shocked, and therefore do not follow a power-law energy distribution, a scenario that has also been suggested in X-ray binaries \citep[e.g.][]{1979ApJ...228..268J,2007ApJ...670..600G,2010MNRAS.405.1759R,2013MNRAS.434.2696S,2018MNRAS.478L.122R}.

A continuum distribution as that shown by NGC\,1052 is $0.3\, \rm{dex}$ above\footnote{For an intrinsic X-ray luminosity of  $\log(L_{2-10\rm{keV}}/\rm{erg\,s^{-1}}) = 41.45 \pm 0.08$, a mid-IR luminosity of $\log(L_{12\micron}/\rm{erg\,s^{-1}}) \sim 41.83$, compared to the $\log(L_{12\micron}/\rm{erg\,s^{-1}}) = 42.12 \pm 0.06$ measured for NGC\,1052.} the mid-IR\,--\,X-ray correlation derived by \citet{2015MNRAS.454..766A}, i.e. over-luminous in the mid-IR although still within the intrinsic scatter. Note that the compact jet scenario would also imply a correlation between IR and X-ray fluxes, since the X-ray emission would rely on the Comptonisation of available IR-to-UV seed photons produced by the jet. The flux distribution shows also a $2500$\,\AA \,--\,$2\, \rm{keV}$ spectral index\footnote{$\alpha_{\rm ox} \sim -0.6$, derived from the high-angular resolution measurements in Table\,\ref{sed_high}.} in agreement with brighter nuclei dominated by a prominent accretion disc \citep{2007MNRAS.377.1696M}. This means that \textit{NGC\,1052 alike nuclei are able to mimic the values typically associated with the presence of tori and accretion discs, despite the lack of a thermal continuum signature in the flux distribution}. The steep continuum explains the elusive nature of the nuclear emission in the IR-to-optical range for NGC\,1052, a characteristic likely extensible to a number of objects that belong to the LLAGN class (e.g. NGC\,1097, M87, Sombrero Galaxy in \citealt{2015ApJ...814..139K}). The existence of a truncated accretion disc is suggested by the detection of a broad Fe\,K$\alpha$ line at $6.4\, \rm{keV}$ \citep{2009ApJ...698..528B}, while the presence of warm dust is indicated by the $9.7\, \rm{\micron}$ silicate emission (Fig.\,\ref{fig_irspec}). Nevertheless, their net contribution to the overall continuum emission in NGC\,1052 appears to be negligible, as self-Compton synchrotron emission from a compact jet seems to dominate the energy output across the whole electromagnetic spectrum, that is over eleven orders of magnitude in frequency.

\section{Summary}\label{sum}

In this work we acquired \textit{N}-band interferometric observations with VLTI/MIDI to investigate the nature of the mid-IR emission in one of the few LLAGN that can be observed using this technique, the prototypical LINER nucleus in NGC\,1052. Additionally, we measure the degree and position angle of the linear polarisation in the near- to mid-IR continuum, using WHT/LIRIS and GTC/CanariCam. With MIDI we found an unresolved nucleus ($< 5\, \rm{mas} \sim 0.5\, \rm{pc}$) up to the longest baselines available to the UTs ($\sim 130\, \rm{m}$; Section\,\ref{results}). We explored two main scenarios for the unresolved emission: an exceptionally compact dust distribution and a compact jet. After a detailed analysis we favour the latter scenario (Section\,\ref{compactjet}), which is further supported by several points of evidence: \textit{i)} the broken power-law shape shown by the continuum spectral distribution from radio to UV wavelengths; \textit{ii)} a $\sim 4\%$ degree of polarisation measured in mid-IR range, together with a low extinction along the line of sight ($A_V \sim 1\, \rm{mag}$); \textit{iii)} the ``harder when brighter'' behaviour in the X-rays indicative of self-Compton synchrotron radiation, a known property of X-ray binaries in the hard state and blazars. Alternatively, if the mid-IR core is dominated by thermal dust emission, this would require first an exceptionally compact dust distribution when compared to those measured in nearby AGN (e.g. Circinus, see Section\,\ref{compactdust}), and second an extinction of $A_V \sim 26\, \rm{mag}$ ($2.8\, \rm{mag}$ assuming the maximum efficiency for the dichroic mechanism). The latter is in contrast with several observational facts: the $A_V \sim 1\, \rm{mag}$ determined from \textit{HST}/FOS observations, the well defined power-law shape of the IR-to-UV continuum, the detection of the silicate band in emission for a highly inclined system, the presence of broad H$\alpha$ and H$\beta$ lines in the unpolarised spectrum, and the increasing hardness at X-rays (higher N$_{\rm H}$) with increasing luminosity.

A distinctive characteristic of NGC\,1052 when compared to jets in classical radio galaxies is the steepness of the power-law continuum in the IR-to-UV range ($\alpha \sim 2.6$). However such a steep distribution could be common among LLAGN, as suggested by the steep ($\alpha \sim 2$) optical/UV power-law continuum observed in LINERs. A similar behaviour has been reported also for X-ray binaries in the low/hard state. The steep IR-to-UV continuum explains the elusive nature of NGC\,1052 --\,and possibly a number of LLAGN\,-- at optical wavelengths. Furthermore, despite the lack of a thermal continuum in the flux distribution, NGC\,1052 does not depart largely from the mid-IR\,--\,X-ray correlation, and also mimics the UV\,--\,X-ray ratios of brighter nuclei. This behaviour is usually driven by a continuum emission dominated by the torus and the accretion disc, respectively. No significant IR variability has been detected for this nucleus over the 2007--2015 period, although a monitoring campaign in the UV would be required to probe the synchrotron radiation variability at the highest energies, also searching for possible occasional bursts as those reported in the millimetre range by \citet{1983ApJ...267L..11H}.

The combination of multi-wavelength analysis, polarimetric, and interferometric observations used in this work offers a unique approach to investigate the nature of the nuclear continuum emission in nearby galaxies. The case of NGC\,1052 suggests that compact jets might be common at the heart of several LLAGN. Future high-angular resolution observations of LLAGN with the \textit{James Webb Space Telescope} (\textit{JWST}) and the Extreme Large Telescopes (ELTs) will be crucial to shed light on the nature of the continuum radiation in these nuclei.

\section*{Acknowledgements}
The authors acknowledge the referee for his/her useful comments that helped to improve the manuscript. This study is based on observations collected at the European Southern Observatory, Chile, programmes 093.B-0616, 094.B-0918, 086.B-0349, 095.B-0376. The following ALMA data has also been used in this work: ADS/JAO.ALMA\#2013.1.00749.S. ALMA is a partnership of ESO (representing its member states), NSF (USA) and NINS (Japan), together with NRC (Canada), MOST and ASIAA (Taiwan), and KASI (Republic of Korea), in cooperation with the Republic of Chile. The Joint ALMA Observatory is operated by ESO, AUI/NRAO and NAOJ. This research has made use of \textsc{astropy} \footnote{\url{http://www.astropy.org}}, a community-developed core \textsc{python} package for Astronomy \citep{2013A&A...558A..33A,2018AJ....156..123A}, and the NASA/IPAC Infrared Science Archive, which is operated by the Jet Propulsion Laboratory, California Institute of Technology, under contract with the National Aeronautics and Space Administration.

JAFO and MAP acknowledge financial support from the Spanish Ministry of Economy under grant number MEC-AYA2015-53753-P. The research leading to these results has received funding from the European Union's Horizon 2020 research and innovation programme under Grant Agreement 730890 (OPTICON). DA acknowledges support from the European Union’s Horizon 2020 Innovation programme under the Marie Sklodowska-Curie grant agreement no. 793499 (DUSTDEVILS).



\bibliographystyle{mnras}
\bibliography{n1052_midi}
\bsp	
\label{lastpage}


\onecolumn
\appendix


\begin{table}

\section{The spectral flux distribution of NGC\,1052}\label{app_sed}

   \caption{Spectral flux distribution for the nucleus of NGC\, 1052 at subarcsec aperture resolution flux. Tabulated values have not been corrected for Galactic reddening. Intrinsic X-ray fluxes are derived from apertures of a few to several arcseconds in size, however the host galaxy contribution in this range is negligible, thus the fluxes are assumed as representative from the nuclear source. The asterisk symbol in the flux column denotes the presence of multiple jet components resolved within the inner $0\farcs15$ in the $2.3$--$89\, \rm{GHz}$ range. In this case the tabulated fluxes correspond to the sum of all the jet components.}\label{sed_high}
   \centering
   \begin{tabular}{r c c c l l}
     \input{tabs/n1052_high.tex}
   \end{tabular}
\end{table}


\begin{table}
   \caption{Spectral flux distribution for the nucleus of NGC\, 1052 extracted from large aperture ($\gtrsim 1''$) flux measurements.}\label{sed_low}
   \centering
   \begin{tabular}{r c c l l}
     \input{tabs/n1052_low.tex}
   \end{tabular}
\end{table}



\begin{table}
\section{Observation log}\label{app_log}

\caption{Log of the VLTI/MIDI campaign for NGC\,1052. The columns in the table correspond to: the time at which the science target fringes were scanned, the total number of frames within the scan (NDIT), the projected baseline length (BL), the baseline position angle (PA), the quality of the interferometric measurement (good: 1, bad: 0), the number of good frames in the scan (NGOOD), the time of the calibrator fringe scan (Caltime), and the difference in airmass between the science target and the calibrator ($\Delta$am).}
\centering
\begin{tabular}{c c c c c c c c}
\multicolumn{4}{c}{Fringe track} & \multicolumn{2}{c}{FT QC} &  \multicolumn{2}{c}{Calibration} \\
Time & NDIT & BL [m] & PA & Q & NGOOD & Caltime & $\Delta$am \\
\hline
\multicolumn{8}{l}{2014-08-06: U1U2} \\
08:08:11 &  8000 &  51 & $10\degr$ &  0 &   303 & 07:49:36 &     -0.1\\
08:15:59 &  8000 &  51 & $11\degr$ &  0 &  1231 & 07:49:36 &     -0.2\\
08:20:52 & 16000 &  51 & $12\degr$ &  0 &   941 & 07:49:36 &     -0.2\\
08:36:02 & 12000 &  51 & $14\degr$ &  0 &  2374 & 08:56:10 &      0.0\\
08:42:13 & 12000 &  51 & $15\degr$ &  1 &   775 & 08:56:10 &      0.0\\
09:26:18 &  8000 &  53 & $21\degr$ &  1 &  2846 & 09:53:16 &      0.0\\
09:31:06 & 12000 &  53 & $21\degr$ &  1 &  8475 & 09:53:16 &      0.0\\
09:36:22 & 12000 &  53 & $22\degr$ &  0 &  7209 & 09:53:16 &      0.0\\
09:42:24 & 12000 &  53 & $23\degr$ &  1 &  3970 & 09:53:16 &      0.0\\
10:14:01 &  8000 &  54 & $26\degr$ &  0 &  2685 & 10:31:19 &      0.0\\
10:17:48 &  8000 &  54 & $26\degr$ &  0 &  5003 & 10:31:19 &      0.0\\
10:21:51 & 12000 &  54 & $27\degr$ &  1 &  8051 & 10:31:19 &      0.0\\
\hline
\multicolumn{8}{l}{2014-08-07: U1U4} \\
09:31:24 &  8000 & 124 & $58\degr$ &  0 &  1632 & 09:20:26 &     -0.0\\
09:35:29 & 12000 & 124 & $58\degr$ &  1 &  3490 & 09:20:26 &     -0.0\\
09:41:18 & 12000 & 125 & $59\degr$ &  1 &  4750 & 09:20:26 &     -0.0\\
10:15:08 &  8000 & 128 & $61\degr$ &  0 &  1938 & 10:03:49 &     -0.0\\
10:19:44 & 12000 & 129 & $61\degr$ &  0 &   230 & 10:03:49 &     -0.0\\
10:28:15 &  8000 & 129 & $62\degr$ &  0 &   345 & 10:03:49 &     -0.0\\
\hline
\multicolumn{8}{l}{2014-11-02: U1U4} \\
02:56:27 & 8000 & 114 & $53\degr$ &  0 &   622 & 02:37:43 &     -0.1\\
03:00:52 & 8000 & 115 & $54\degr$ &  0 &   557 & 02:37:43 &     -0.1\\
03:05:35 & 8000 & 116 & $54\degr$ &  0 &   931 & 02:37:43 &     -0.1\\
03:37:51 & 8000 & 122 & $57\degr$ &  0 &  1323 & 03:54:07 &      0.0\\
03:42:02 & 8000 & 123 & $58\degr$ &  0 &  1218 & 03:54:07 &      0.0\\
03:45:51 & 8000 & 123 & $58\degr$ &  0 &  1062 & 03:54:07 &     -0.0\\
04:08:09 & 8000 & 126 & $59\degr$ &  0 &   361 & 03:54:07 &     -0.0\\
04:11:58 & 8000 & 127 & $60\degr$ &  0 &   325 & 03:54:07 &     -0.0\\
04:16:05 & 8000 & 127 & $60\degr$ &  0 &  1007 & 03:54:07 &     -0.0\\
04:19:54 & 8000 & 127 & $60\degr$ &  0 &   196 & 03:54:07 &     -0.0\\
04:23:51 & 8000 & 128 & $60\degr$ &  0 &   624 & 03:54:07 &     -0.0\\
\hline
\multicolumn{8}{l}{2014-11-03: U2U3} \\
03:05:30 & 8000 &  42 & $32\degr$ &  0 &   796 & 02:22:14 &     -0.1\\
\hline
\multicolumn{8}{l}{2014-11-03: U3U4} \\
05:12:34 & 8000 &  60 & $112\degr$ &  0 &  1146 & 05:21:24 &      0.0\\
05:36:56 & 8000 &  58 & $114\degr$ &  0 &   411 & 05:21:24 &      0.0\\
05:40:53 & 8000 &  57 & $114\degr$ &  0 &   457 & 05:21:24 &      0.0\\
\hline
\multicolumn{8}{l}{2014-11-04: U3U4} \\
07:30:09 & 8000 &  43 & $128\degr$ &  0 &  4053 & 07:12:53 &      0.2\\
07:33:38 & 8000 &  42 & $129\degr$ &  1 &  4736 & 07:12:53 &      0.2\\
07:37:12 & 8000 &  42 & $130\degr$ &  1 &  5039 & 07:12:53 &      0.2\\
07:58:36 & 8000 &  38 & $135\degr$ &  0 &   193 & 08:14:20 &      0.0\\
08:02:25 & 8000 &  38 & $136\degr$ &  1 &  3981 & 08:14:20 &      0.0\\
08:06:11 & 8000 &  37 & $137\degr$ &  1 &  5169 & 08:14:20 &      0.1\\
08:24:55 & 8000 &  35 & $142\degr$ &  0 &   253 & 08:14:20 &      0.3\\
08:28:23 & 8000 &  34 & $143\degr$ &  0 &   225 & 08:14:20 &      0.3\\
08:32:15 & 8000 &  34 & $145\degr$ &  0 &  1677 & 08:14:20 &      0.4\\
\hline
\end{tabular}
\end{table}


\end{document}

%% file: tabs/n1052_high.tex
%
\bf Frequency [Hz] & \bf F$_\nu$ [Jy] & \bf Aperture [radius/beam] & \bf Telescope/Band & \bf Reference \\
\hline
$   2.53 \times 10^{19}$ & $(1.5 \pm 0.3) \times 10^{-7}$ &                   & \textit{Swift/BAT}, $14$--$195\, \rm{keV}$ & \citealt{2010ApJS..186..378T} \\[0.05cm]
$   1.99 \times 10^{19}$ & $(1.4 \pm 0.8) \times 10^{-7}$ &                   & \textit{Swift/BAT}, $15$--$150\, \rm{keV}$ & {\citealt{2010A&A...524A..64C}} \\[0.05cm]
$ 1.6926 \times 10^{19}$ & $1.1 \times 10^{-7}$ &                             & \textit{INTEGRAL}, $40$--$100\, \rm{keV}$  & {\citealt{2009A&A...505..417B}} \\[0.05cm]
$  8.463 \times 10^{18}$ & $3.0 \times 10^{-7}$ &                             & \textit{Suzaku}, $10$--$60\, \rm{keV}$     & \citealt{2009ApJ...698..528B} \\[0.05cm]
$  8.463 \times 10^{18}$ & $(1.6 \pm 0.2) \times 10^{-7}$ &                   & \textit{Swift/BAT}, $15$--$55\, \rm{keV}$  & \citealt{2009ApJ...699..603A} \\[0.05cm]
$  7.254 \times 10^{18}$ & $1.7 \times 10^{-7}$ &                             & \textit{INTEGRAL}, $20$--$40\, \rm{keV}$   & {\citealt{2009A&A...505..417B}} \\[0.05cm]
$ 1.4508 \times 10^{18}$ & $2.9^{+0.75}_{-2.44} \times 10^{-7}$ &                & \textit{Chandra}, $2$--$10\, \rm{keV}$     & {\citealt{2009A&A...506.1107G}} \\[0.05cm]
$ 1.4508 \times 10^{18}$ & $4.6 \times 10^{-7}$ &                             & \textit{Suzaku}, $2$--$10\, \rm{keV}$      & \citealt{2009ApJ...698..528B} \\[0.05cm]
$ 1.1046 \times 10^{15}$ & $(8.1 \pm 0.3) \times 10^{-6}$ & $0\farcs15$       & \textit{HST}/ACS-HRC, F250W    & This work     \\[0.05cm]
$ 8.9206 \times 10^{14}$ & $(1.48 \pm 0.03) \times 10^{-5}$ & $0\farcs15$     & \textit{HST}/ACS-HRC, F330W    & This work     \\[0.05cm]
$ 5.5006 \times 10^{14}$ & $(3.4 \pm 0.1) \times 10^{-5}$ & $0\farcs15$       & \textit{HST}/WFPC1, F555W      & This work     \\[0.05cm]
$ 2.3715 \times 10^{14}$ & $(6.3 \pm 0.1) \times 10^{-4}$ & $0\farcs26$       & VLT/NaCo, \textit{J} band      & This work     \\[0.05cm]
$  1.868 \times 10^{14}$ & $(8.0 \pm 0.3) \times 10^{-4}$ & $0\farcs15$       & \textit{HST}/NIC2, F160W       & This work     \\[0.05cm]
$ 1.3761 \times 10^{14}$ & $(1.19 \pm 0.02) \times 10^{-3}$ & $0\farcs15$     & VLT/NaCo, \textit{Ks} band     & This work     \\[0.05cm]
$ 7.8947 \times 10^{13}$ & $(7.17 \pm 0.03) \times 10^{-3}$ & $0\farcs15$     & VLT/NaCo, \textit{L'} band     & This work     \\[0.05cm]
$   3.86 \times 10^{13}$ & $(58.8 \pm 9.4) \times 10^{-3}$ & $0\farcs52 \times 0\farcs39$      & VLT/VISIR, J7.9          & \citealt{2014MNRAS.439.1648A} \\[0.05cm]
$   3.49 \times 10^{13}$ & $(62.4 \pm 14.4) \times 10^{-3}$ & $0\farcs45 \times 0\farcs39$      & VLT/VISIR, PAH1          & \citealt{2014MNRAS.439.1648A} \\[0.05cm]
$   3.49 \times 10^{13}$ & $(58 \pm 14) \times 10^{-3}$   & $0\farcs45 \times 0\farcs40$      & VLT/VISIR, PAH1          & This work \\[0.05cm]
$  3.446 \times 10^{13}$ & $(63 \pm 6) \times 10^{-3}$    &   $0\farcs5$                      & GTC/CanariCam, Si2                             & This Work \\[0.03cm]
$   3.43 \times 10^{13}$ & $(63.6 \pm 1.6) \times 10^{-3}$   & $0\farcs75 \times 0\farcs62$      & Gemini/T-ReCS, Si2       & \citealt{2014MNRAS.439.1648A} \\[0.05cm]
$   3.05 \times 10^{13}$ & $(114.7 \pm 22.9) \times 10^{-3}$ & $0\farcs39 \times 0\farcs35$     & VLT/VISIR, B9.7          & \citealt{2014MNRAS.439.1648A} \\[0.05cm]
$   2.81 \times 10^{13}$ & $(133.7 \pm 22.8) \times 10^{-3}$ & $0\farcs43 \times 0\farcs37$     & VLT/VISIR, B10.7         & \citealt{2014MNRAS.439.1648A} \\[0.05cm]
$    2.6 \times 10^{13}$ & $(154.1 \pm 13.2) \times 10^{-3}$ & $0\farcs44 \times 0\farcs40$     & VLT/VISIR, B11.7         & \citealt{2014MNRAS.439.1648A} \\[0.05cm]
$  2.584 \times 10^{13}$ & $(123 \pm 12) \times 10^{-3}$ & $0\farcs5$                          & GTC/CanariCam, Si5                             & This Work \\[0.03cm]
$   2.52 \times 10^{13}$ & $(133.9 \pm 4.2) \times 10^{-3}$ & $0\farcs40 \times 0\farcs36$     & VLT/VISIR, PAH2\_2       & \citealt{2014MNRAS.439.1648A} \\[0.05cm]
$   2.52 \times 10^{13}$ & $(133 \pm 4) \times 10^{-3}$   & $0\farcs40 \times 0\farcs40$     & VLT/VISIR, PAH2\_2       & This work \\[0.05cm]
$    2.4 \times 10^{13}$ & $(181.2 \pm 19.8) \times 10^{-3}$ & $0\farcs51 \times 0\farcs42$     & VLT/VISIR, B12.4         & \citealt{2014MNRAS.439.1648A} \\[0.05cm]
$   1.64 \times 10^{13}$ & $(355.7 \pm 9.1) \times 10^{-3}$ & $0\farcs55 \times 0\farcs52$     & Gemini/T-ReCS, Qa        & \citealt{2014MNRAS.439.1648A} \\[0.05cm]
$    1.6 \times 10^{13}$ & $(339.0 \pm 33.8) \times 10^{-3}$ & $0\farcs50 \times 0\farcs49$     & VLT/VISIR, Q2            & \citealt{2014MNRAS.439.1648A} \\[0.05cm]
$  4.778 \times 10^{11}$ & $(201 \pm 2) \times 10^{-3}$      & $0\farcs12 \times 0\farcs10$     & ALMA, $478\, \rm{GHz}$   & This work \\[0.05cm]
$  4.657 \times 10^{11}$ & $(207 \pm 1) \times 10^{-3}$      & $0\farcs12 \times 0\farcs10$     & ALMA, $466\, \rm{GHz}$   & This work \\[0.05cm]
$  4.315 \times 10^{11}$ & $(273 \pm 1) \times 10^{-3}$      & $0\farcs13 \times 0\farcs12$     & ALMA, $432\, \rm{GHz}$   & This work \\[0.05cm]
$  4.194 \times 10^{11}$ & $(283 \pm 1) \times 10^{-3}$      & $0\farcs14 \times 0\farcs12$     & ALMA, $419\, \rm{GHz}$   & This work \\[0.05cm]
$  3.528 \times 10^{11}$ & $(283.1 \pm 1.1) \times 10^{-3}$  & $0\farcs20 \times 0\farcs15$     & ALMA, $353\, \rm{GHz}$   & \citealt{2019arXiv190102280P} \\[0.05cm]
$  3.408 \times 10^{11}$ & $(299.3 \pm 1.1) \times 10^{-3}$  & $0\farcs20 \times 0\farcs15$     & ALMA, $341\, \rm{GHz}$   & \citealt{2019arXiv190102280P} \\[0.05cm]
$  2.294 \times 10^{11}$ & $(457 \pm 4) \times 10^{-3}$      & $0\farcs3 \times 0\farcs2$       & ALMA, $229\, \rm{GHz}$   & \citealt{2019arXiv190102280P} \\[0.05cm]
$  2.141 \times 10^{11}$ & $(495 \pm 3) \times 10^{-3}$      & $0\farcs3 \times 0\farcs2$       & ALMA, $214\, \rm{GHz}$   & \citealt{2019arXiv190102280P} \\[0.05cm]
$  1.080 \times 10^{11}$ & $(750 \pm 7) \times 10^{-3}$      & $0\farcs7 \times 0\farcs6$       & ALMA, $108\, \rm{GHz}$   & \citealt{2019arXiv190102280P} \\[0.05cm]
$    8.9 \times 10^{10}$ & $0.561^*$           & $1.44 \times 0.90\, \rm{mas^2}$ & KVN, $89\, \rm{GHz}$                    & \citealt{2016ApJ...830L...3S} \\[0.05cm]
$    8.6 \times 10^{10}$ & $0.62 \pm 0.09^*$   & $0.407 \times 0.075\, \rm{mas^2}$ & GMVA, $86\, \rm{GHz}$  & {\citealt{2016A&A...593A..47B}} \\[0.05cm]
$    4.3 \times 10^{10}$ & $0.49$--$1.64$            & $1.11  \times 0.21\, \rm{mas^2}$  & VLBI, $43\, \rm{GHz}$, min--max 2005-2009    & \citealt{2019arXiv190102639B} \\[0.05cm]
$   2.21 \times 10^{10}$ & $2.46 \pm 0.02$   & $0\farcs09 \times 0\farcs07$       & JVLA, $22\, \rm{GHz}$   & \citealt{2019arXiv190102280P} \\[0.05cm]
$    2.2 \times 10^{10}$ & $0.50$--$2.26$            & $0.96  \times 0.37\, \rm{mas^2}$  & VLBI, $22\, \rm{GHz}$, min--max 2005-2009    & \citealt{2019arXiv190102639B} \\[0.05cm]
$ 1.5364 \times 10^{10}$ & $1.99 \pm 0.04^*$ & $1.03  \times 0.40\, \rm{mas^2}$  & VLBA, $15\, \rm{GHz}$ & \citealt{2001PASJ...53..169K} \\[0.05cm]
$   1.49 \times 10^{10}$ & $1.0 \pm 0.1$     & $0\farcs1 \times 0\farcs09$       & JVLA, $15\, \rm{GHz}$   & \citealt{2019arXiv190102280P} \\[0.05cm]
$    8.4 \times 10^{9}$ & $2.91 \pm 0.03$     & $0\farcs3 \times 0\farcs2$       & JVLA, $8.4\, \rm{GHz}$   & \citealt{2019arXiv190102280P} \\[0.05cm]
$    8.4 \times 10^{9}$ & $2.39^*$           & $1.98  \times 0.81\, \rm{mas^2}$  & VLBI, $8.4\, \rm{GHz}$                  & {\citealt{2004A&A...426..481K}} \\[0.05cm]
$    6.2 \times 10^{9}$ & $1.295 \pm 0.001$     & $0\farcs4 \times 0\farcs3$     & JVLA, $6.2\, \rm{GHz}$   & \citealt{2019arXiv190102280P} \\[0.05cm]
$    5.0 \times 10^{9}$ & $2.41^*$           & $3.3   \times 1.31\, \rm{mas^2}$  & VLBI, $5\, \rm{GHz}$                    & {\citealt{2004A&A...426..481K}} \\[0.05cm]
$   2.32 \times 10^{9}$ & $1.51^*$           & $9.0   \times 3.8\, \rm{mas^2}$   & VLBA, $2.3\, \rm{GHz}$                  & \citealt{1997ApJS..111...95F} \\[0.05cm]
$ 1.6710 \times 10^{9}$ & $0.62 \pm 0.06$  & $3.40\, \rm{mas}$               & VLBI, $1.7\, \rm{GHz}$                      & \citealt{1979ApJ...233L.105S} \\[0.05cm]
$   3.27 \times 10^{8}$ & $0.30 \pm 0.15$   & $0\farcs1$                        & Ooty Radio Telescope, $327\, \rm{MHz}$    & \citealt{1984IAUS..110..261B} \\[0.05cm]
\hline

%% file: tabs/n1052_low.tex
%
%
\bf Frequency & \bf F$_\nu$ &      \bf Aperture & \bf Telescope/Band                              & \bf Reference                            \\ 
       {[Hz]} &       [Jy]  &     [radius/beam] &                                                &                                       \\ 
\hline
$   1.95 \times 10^{15}$ &$ (7.7 \pm 0.6) \times 10^{-5}$ & $4\farcs21$ & \textit{GALEX} FUV                             & \citealt{2011MNRAS.411..311M} \\[0.03cm]
$   1.29 \times 10^{15}$ &$ (1.47 \pm 0.06) \times 10^{-4}$ & $4\farcs21$ & \textit{GALEX} NUV                             & \citealt{2011MNRAS.411..311M} \\[0.03cm]
$   2.38 \times 10^{14}$ &$ (2.6 \pm 0.3) \times 10^{-2}$ &       $2''$ & InfraRed Telescope Facility (IRTF), \textit{J} band & \citealt{1982ApJ...263..624B} \\[0.03cm]
$   1.87 \times 10^{14}$ &$ (3.5 \pm 0.3) \times 10^{-2}$ &       $2''$ & IRTF, \textit{H} band                          & \citealt{1982ApJ...263..624B} \\[0.03cm] 
$   1.35 \times 10^{14}$ &$ (3.2 \pm 0.3) \times 10^{-2}$ &       $2''$ & IRTF, \textit{K} band                          & \citealt{1982ApJ...263..624B} \\[0.03cm]
$   8.47 \times 10^{13}$ &$ (2.8 \pm 0.3) \times 10^{-2}$ &       $2''$ & IRTF, \textit{L} band                          & \citealt{1982ApJ...263..624B} \\[0.03cm]
$ 8.3333 \times 10^{13}$ &$ 4.0 \times 10^{-2}$ &  $2\farcs4$ & \textit{Spitzer} $3.6\, \rm{\micron}$          & This Work                   \\[0.03cm] 
$ 6.6667 \times 10^{13}$ &$ 3.7 \times 10^{-2}$ &  $2\farcs4$ & \textit{Spitzer} $4.5\, \rm{\micron}$          & This Work                   \\[0.03cm] 
$   6.25 \times 10^{13}$ &$ (3.0 \pm 0.4) \times 10^{-2}$ &       $4''$ & IRTF, \textit{M} band                          & \citealt{1982ApJ...263..624B} \\[0.03cm]
$ 5.1724 \times 10^{13}$ &$ 4.7 \times 10^{-2}$ &  $2\farcs4$ & \textit{Spitzer} $5.8\, \rm{\micron}$          & This Work                                       \\[0.03cm] 
$   3.75 \times 10^{13}$ &$ 6.9 \times 10^{-2}$ &  $2\farcs4$ & \textit{Spitzer} $8.0\, \rm{\micron}$          & This Work                                       \\[0.03cm] 
$ 3.3333 \times 10^{13}$ & $(15 \pm 3) \times 10^{-2}$ & $\sim 5\farcs5$ & \textit{Akari}/IRC $9\, \rm{\micron}$, Point Source Catalogue & NASA/IPAC Infrared Science Archive \\[0.03cm] 
$   2.83 \times 10^{13}$ & $(11.1 \pm 1.6) \times 10^{-2}$ & $4''$ & IRTF, \textit{N} band                          & \citealt{1982ApJ...263..624B}  \\[0.03cm] 
$  2.595 \times 10^{13}$ & $(15.7 \pm 0.2) \times 10^{-2}$ &  $7\farcs4$ & \textit{WISE} W3, AllWISE Source Catalog       & NASA/IPAC Infrared Science Archive             \\[0.03cm] 
$    2.5 \times 10^{13}$ & $(13.8 \pm 0.2) \times 10^{-2}$ & $\sim 30''$ & \textit{IRAS} $12\, \rm{\micron}$, IRS enhanced products & NASA/IPAC Infrared Science Archive \\[0.03cm]
$ 1.6667 \times 10^{13}$ & $(38 \pm 0.02) \times 10^{-2}$ & $\sim 5\farcs7$ & \textit{Akari}/IRC $18\, \rm{\micron}$, Point Source Catalogue & NASA/IPAC Infrared Science Archive \\[0.03cm]
$   1.47 \times 10^{13}$ & $(46 \pm 7) \times 10^{-2}$ &   $4''$ & IRTF, \textit{Q} band                          & \citealt{1982ApJ...263..624B}  \\[0.03cm]
$ 1.3582 \times 10^{13}$ & $(44.3 \pm 0.9) \times 10^{-2}$ & $\sim 12''$ & \textit{WISE} W4, AllWISE Source Catalog       & NASA/IPAC Infrared Science Archive             \\[0.03cm] 
$   1.27 \times 10^{13}$ & $(41.4 \pm 0.8) \times 10^{-2}$ &     & \textit{Spitzer}/MIPS $24\, \rm{\micron}$, IRS enhanced products & NASA/IPAC Infrared Science Archive \\[0.03cm]
$    1.2 \times 10^{13}$ & $(42.2 \pm 0.7) \times 10^{-2}$ & $\sim 30''$ & \textit{IRAS} $25\, \rm{\micron}$, IRS enhanced products & NASA/IPAC Infrared Science Archive \\[0.03cm]
$      5 \times 10^{12}$ &                 $0.94$ & $\sim 60''$ & \textit{IRAS} $60\, \rm{\micron}$, Point Source Catalog v2.1 & NASA/IPAC Infrared Science Archive \\[0.03cm]
$ 4.6154 \times 10^{12}$ &                $0.84$ & $\sim  50''$ & \textit{Akari}/FIS $65\, \rm{\micron}$, Bright Source Catalogue & NASA/IPAC Infrared Science Archive             \\[0.03cm]
$ 4.2827 \times 10^{12}$ & $(73.7 \pm 1.5) \times 10^{-2}$ & $6\farcs25 \times 5\farcs35$ & \textit{Herschel}/PACS $70\, \rm{\micron}$, Point Source Catalog & NASA/IPAC Infrared Science Archive \\[0.03cm]
$ 3.3333 \times 10^{12}$ & $(79 \pm 4) \times 10^{-2}$ & $\sim  50''$ & \textit{Akari}/FIS $90\, \rm{\micron}$, Bright Source Catalogue & NASA/IPAC Infrared Science Archive             \\[0.03cm]
$      3 \times 10^{12}$ &                 $1.22$ &  $\sim 2'$ & \textit{IRAS} $100\, \rm{\micron}$, Point Source Catalog v2.1 & NASA/IPAC Infrared Science Archive \\[0.03cm]
$ 2.1429 \times 10^{12}$ &          $1.3 \pm 0.4$ & $\sim  90''$ & \textit{Akari}/FIS $140\, \rm{\micron}$, Bright Source Catalogue & NASA/IPAC Infrared Science Archive             \\[0.03cm]
$  1.875 \times 10^{12}$ &                  $0.9$ & $\sim  90''$ & \textit{Akari}/FIS $160\, \rm{\micron}$, Bright Source Catalogue & NASA/IPAC Infrared Science Archive             \\[0.03cm]
$ 1.8737 \times 10^{12}$ & $0.65 \pm 0.18$ & $17\farcs88 \times 11\farcs52$ & \textit{Herschel}/PACS $160\, \rm{\micron}$, Point Source Catalog & NASA/IPAC Infrared Science Archive \\[0.03cm]
$  6.667 \times 10^{11}$ &      $0.941 \pm 0.785$ &      $15''$ & JCMT $0.45\, \rm{mm}$                          & \citealt{1991AJ....101.1609K} \\[0.03cm] 
$   3.75 \times 10^{11}$ &      $0.412 \pm 0.039$ &      $17''$ & JCMT $0.8\, \rm{mm}$                           & \citealt{1991AJ....101.1609K} \\[0.03cm]
$ 2.7273 \times 10^{11}$ &      $0.511 \pm 0.019$ & $18\farcs5$ & JCMT $1.1\, \rm{mm}$                           & \citealt{1991AJ....101.1609K} \\[0.03cm]
$ 2.3077 \times 10^{11}$ &      $0.563 \pm 0.028$ & $19\farcs5$ & JCMT $1.3\, \rm{mm}$                           & \citealt{1991AJ....101.1609K} \\[0.03cm]
$    1.5 \times 10^{11}$ &      $1.036 \pm 0.083$ & $27\farcs5$ & JCMT $2.0\, \rm{mm}$                           & \citealt{1991AJ....101.1609K} \\[0.03cm]
$   8.96 \times 10^{10}$ &          $2.0 \pm 0.5$ & $\sim 60''$ & NRAO 11\,m telescope $89.6\, \rm{GHz}$, event A 05/1973   & \citealt{1983ApJ...267L..11H} \\[0.03cm]
$   8.96 \times 10^{10}$ &          $1.2 \pm 0.5$ & $\sim 60''$ & NRAO 11\,m telescope $89.6\, \rm{GHz}$, event B 02/1980   & \citealt{1983ApJ...267L..11H} \\[0.03cm]
$   8.57 \times 10^{10}$ &        $0.91 \pm 0.24$ & $\sim 65''$ & NRAO 11\,m telescope $3.5\, \rm{mm}$, average 1968-1973   & \citealt{1975ApJ...196..347H} \\[0.03cm]
$   8.57 \times 10^{10}$ &        $2.21 \pm 0.52$ & $\sim 65''$ & NRAO 11\,m telescope $3.5\, \rm{mm}$, 07/1973             & \citealt{1975ApJ...196..347H} \\[0.03cm]
$   3.16 \times 10^{10}$ &        $1.52 \pm 0.24$ &   $\sim 3'$ & NRAO 11\,m telescope $9.5\, \rm{mm}$, average 1968-1973   & \citealt{1975ApJ...196..347H} \\[0.03cm]
$   3.16 \times 10^{10}$ &        $2.86 \pm 0.24$ &   $\sim 3'$ & NRAO 11\,m telescope $9.5\, \rm{mm}$, 07/1973             & \citealt{1975ApJ...196..347H} \\[0.03cm]
$   3.14 \times 10^{10}$ &          $2.9 \pm 0.2$ &   $\sim 3'$ & NRAO 11\,m telescope $31.4\, \rm{GHz}$, event A 05/1973   & \citealt{1983ApJ...267L..11H} \\[0.03cm]
$   3.14 \times 10^{10}$ &          $1.7 \pm 0.2$ &   $\sim 3'$ & NRAO 11\,m telescope $31.4\, \rm{GHz}$, event B 02/1980   & \citealt{1983ApJ...267L..11H} \\[0.03cm]
$    2.2 \times 10^{10}$ &       $1.62 \pm 0.162$ &       $7''$ & ATCA $22\, \rm{GHz}$                                      & {\citealt{2006A&A...445..465R}} \\[0.03cm]
$   1.55 \times 10^{10}$ &          $2.0 \pm 0.1$ &   $\sim 6'$ & NRAO 11\,m telescope $15.5\, \rm{GHz}$, event A 05/1973   & \citealt{1983ApJ...267L..11H} \\[0.03cm]
$   1.55 \times 10^{10}$ &          $1.8 \pm 0.1$ &   $\sim 6'$ & NRAO 11\,m telescope $15.5\, \rm{GHz}$, event B 02/1980   & \citealt{1983ApJ...267L..11H} \\[0.03cm]
 $    8.6 \times 10^{9}$ &                 $2.63$ &       $1''$ & ATCA $8.6\, \rm{GHz}$                          & \citealt{2003PASJ...55..351T} \\[0.03cm]
 $    4.8 \times 10^{9}$ &                 $2.47$ &       $2''$ & ATCA $4.8\, \rm{GHz}$                          & \citealt{2003PASJ...55..351T} \\[0.03cm]
 $    2.5 \times 10^{9}$ &                 $1.46$ &  $4\farcs5$ & ATCA $2.5\, \rm{GHz}$                          & \citealt{2003PASJ...55..351T} \\[0.03cm]
 $ 1.6649 \times 10^{9}$ & $(83.1 \pm 1.7) \times 10^{-2}$ & $1''$ & VLA $1.66\, \rm{GHz}$                          & \citealt{1984ApJ...284..531W} \\[0.03cm]
 $ 1.3851 \times 10^{9}$ & $(73.5 \pm 1.5) \times 10^{-2}$ & $1''$ & VLA $1.39\, \rm{GHz}$                          & \citealt{1984ApJ...284..531W} \\[0.03cm]
 $   4.08 \times 10^{8}$ &        $0.84 \pm 0.04$ &             & Molonglo Radio Telescope $408\, \rm{MHz}$      & \citealt{1981MNRAS.194..693L} \\[0.03cm]
 $   3.65 \times 10^{8}$ &                 $0.64$ &             & Texas Interferometer $365\, \rm{MHz}$          & \citealt{1996AJ....111.1945D} \\[0.03cm]
 $   7.38 \times 10^{7}$ &        $0.85 \pm 0.13$ & $\sim 80''$ & VLA $74\, \rm{MHz}$                            & \citealt{2007AJ....134.1245C} \\
\hline